%% file: BetaDecaysFCCee.tex
\documentclass[11pt,a4paper]{article}
\pdfoutput=1

\usepackage[colorlinks=true, linkcolor=black!50!blue, urlcolor=blue, citecolor=blue, anchorcolor=blue]{hyperref}
\usepackage[font=small,labelfont=bf,margin=0mm,labelsep=period,tableposition=top]{caption}
\usepackage[a4paper,top=1.5cm,bottom=2cm,left=1.5cm,right=1.5cm,bindingoffset=0mm]{geometry}

\usepackage{placeins,cite}
\usepackage{graphicx}
\usepackage{soul}
\usepackage{subcaption}
\usepackage{float}
\usepackage{afterpage}
\usepackage{epsfig}
\usepackage{xspace}
\usepackage{amssymb}
\usepackage{amsmath}
\usepackage{bm}
\usepackage{multirow}
\usepackage{url}
\usepackage{xcolor}
\usepackage[normalem]{ulem}
\usepackage{url}
\usepackage{booktabs,multirow}
\usepackage{textcomp}
\usepackage{comment}
\usepackage{cleveref}
\graphicspath{{../}{./figures/}}

\usepackage{makecell}

\makeatletter
\def\thickhline{%
             \noalign{\ifnum0 =`}\fi\hrule \@height \thickarrayrulewidth \futurelet
             \reserved@a\@xthickhline}
\def\@xthickhline{\ifx\reserved@a\thickhline
                \vskip\doublerulesep
                \vskip -\thickarrayrulewidth
                \fi
                \ifnum0 =`{\fi}}
\makeatother
\newlength{\thickarrayrulewidth}
\usepackage{tikz}
\usetikzlibrary{positioning}
\usetikzlibrary{shapes.geometric, arrows}
\usetikzlibrary{arrows.meta}
\usepackage{varwidth}
\usepackage{xcolor}
\definecolor{mtplotlib1}{HTML}{1f77b4}
\definecolor{mtplotlib2}{HTML}{ff7f0e}
\definecolor{mtplotlib3}{HTML}{2ca02c}
\definecolor{mtplotlib4}{HTML}{d62728}

\tikzset{%
  >={Latex[width=2mm,length=2mm]},
            base/.style = {rectangle, rounded corners, draw=black,
                           minimum width=4cm, minimum height=1cm,
                           text centered}, 
            mystyle/.style={rectangle, rounded corners, draw=black,
            minimum width=12cm, minimum height=1cm,
            text centered}, 
    col0/.style = {base, fill=white!30},
    col1/.style = {base, fill=mtplotlib1!30},
    col11/.style = {mystyle, fill=mtplotlib1!30},
    col2/.style = {base, fill=mtplotlib2!30},
    col3/.style = {base, fill=mtplotlib3!30},
    col4/.style = {base, minimum width=2.5cm, fill=mtplotlib4!15,}
}

\newcommand{\smefit}{\textsc{SMEFiT}\xspace}

\usepackage[dvipsnames]{xcolor}

\usepackage{makecell}

\newcommand{\lra}[1]{\overleftrightarrow{#1}}

\newcommand{\be}{\begin{equation}}
\newcommand{\ee}{\end{equation}}
\newcommand{\bea}{\begin{eqnarray}}
\newcommand{\eea}{\end{eqnarray}}
\newcommand{\bi}{\begin{itemize}}
\newcommand{\ei}{\end{itemize}}
\newcommand{\ben}{\begin{enumerate}}
\newcommand{\een}{\end{enumerate}}

\newcommand{\lp}{\left(}
\newcommand{\rp}{\right)}

\def\frac#1#2{{{#1}\over {#2}}}
\def\gsim{\mathrel{\rlap{\lower4pt\hbox{\hskip1pt$\sim$}}
    \raise1pt\hbox{$>$}}}       
\def\lsim{\mathrel{\rlap{\lower4pt\hbox{\hskip1pt$\sim$}}
    \raise1pt\hbox{$<$}}}

\newcommand{\draft}[1]{}
\newcommand{\OO}{\ensuremath{\mathcal{O}}}
\newcommand{\Op}[1]{\OO_{\sss #1}}
\newcommand{\sss}{\scriptscriptstyle}

\newcommand{\qq}[3]{\ensuremath{\mathcal{O}_{#2}^{#1 (#3)}}}
\def\beq{\begin{equation}}
\def\eeq{\end{equation}}

\numberwithin{equation}{section}
\numberwithin{figure}{section}
\numberwithin{table}{section}

\newcommand{\GFexp}{G_F^\text{(\rm exp)}}
\newcommand{\wilson}{\texttt{wilson} }
\newcommand{\rgevolve}{\texttt{rgevolve}}

\usepackage{tabularx}
\newcolumntype{C}[1]{>{\centering\arraybackslash}p{#1}}

\definecolor{darkblue}{rgb}{0.0,0,0.5}
\definecolor{darkgreen}{rgb}{0.0,0.3,0.0}
\definecolor{redish}{rgb}{0.675,0,0.2}
\definecolor{red}{rgb}{0.8,0,0}
\definecolor{green}{rgb}{0,0.6,0}
\definecolor{bluish}{rgb}{0.2,0.2,0.675}
\definecolor{mygrey}{rgb}{0.6,0.6,0.6}

\usepackage{tikz} 
\usetikzlibrary{shapes,arrows,positioning,automata,backgrounds,calc,er,patterns,arrows.meta}
\usepackage{tikz-feynman}
\tikzfeynmanset{compat=1.0.0}
\usepackage{varwidth}
\usepackage{xcolor}
\definecolor{mtplotlib1}{HTML}{1f77b4}
\definecolor{mtplotlib2}{HTML}{ff7f0e}
\definecolor{mtplotlib3}{HTML}{2ca02c}
\definecolor{mtplotlib4}{HTML}{d62728}

\tikzset{%
  >={Latex[width=2mm,length=2mm]},
            base/.style = {rectangle, rounded corners, draw=black,
                           minimum width=4cm, minimum height=1cm,
                           text centered}, 
            mystyle/.style={rectangle, rounded corners, draw=black,
            minimum width=12cm, minimum height=1cm,
            text centered}, 
    col0/.style = {base, fill=white!30},
    col1/.style = {base, fill=mtplotlib1!30},
    col11/.style = {mystyle, fill=mtplotlib1!30},
    col2/.style = {base, fill=mtplotlib2!30},
    col3/.style = {base, fill=mtplotlib3!30},
    col4/.style = {base, minimum width=2.5cm, fill=mtplotlib4!15,}
}

\usepackage{tabularx}
\usepackage{subcaption}
\newcolumntype{C}[1]{>{\centering\arraybackslash}p{#1}}

\definecolor{lightblue}{rgb}{0.0,0.5,1.0}

\begin{document}
\newgeometry{top=1.5cm,bottom=1.5cm,left=1.5cm,right=1.5cm,bindingoffset=0mm}

\begingroup\raggedleft 
\hfil{INT-PUB-26-034} \\
\endgroup

$\,\qquad$
\vspace{1.6cm}

\begin{center}
  {\Large \bf Disentangling $V_{ud}$ from 
  New Physics in Beta Decays with the FCC-ee}\\
  \vspace{1.1cm}
  {
 Lemonia Gialidi$^{1,2}$, Elie Hammou$^1$, Kamil Laurent$^{1,3}$, Vaisakh Plakkot$^{1,2,4}$,\\[+0.1cm] Juan Rojo$^{1,3}$, and Jordy de Vries$^{1,2}$
  }\\
  
\vspace{0.7cm}
{\it ~$^{1}$Nikhef Theory Group, Science Park 105, 1098 XG Amsterdam, The Netherlands\\[0.1cm]
~$^{2}$Institute of Physics, Universiteit van Amsterdam, \\Science Park 904, 1098 XH Amsterdam, The Netherlands\\[0.1cm]
~$^{3}$Department of Physics and Astronomy, Vrije Universiteit Amsterdam, \\NL-1081 HV Amsterdam, The Netherlands\\[0.1cm]
~$^{4}$Institute for Nuclear Theory, University of Washington, Seattle, WA 98195, USA\\[0.1cm]}

\vspace{1.0cm}

{\bf \large Abstract}

\end{center}

Superallowed nuclear $\beta$-decays enable the most precise determination of the CKM element $V_{ud}$, but they are only sensitive to an effective combination of $V_{ud}$ and New Physics contributions to left-handed charged currents, which low-energy data alone cannot disentangle.
Here we determine $V_{ud}$ simultaneously with the relevant Wilson coefficients of the Standard Model Effective Field Theory (SMEFT) by combining, within the {\sc\small SMEFiT} framework, a dedicated likelihood  for superallowed transitions, including radiative and nuclear-structure uncertainties, with LEP and LHC world data and  HL-LHC and FCC-ee projections. 
Current collider data and HL-LHC projections only partially lift the degeneracy between $V_{ud}$ and possible New Physics effects. 
Instead, the FCC-ee fully decouples $V_{ud}$ from New Physics, recovering SM-level precision while significantly tightening the constraints on the SMEFT Wilson coefficients. 
This conclusion is robust against variations in flavour assumptions and theory uncertainties. 
Our results highlight the complementarity between future colliders and low-energy experiments, and pave the way towards a SMEFT-consistent interpretation of first-row CKM unitarity tests and the Cabibbo angle anomaly.

\clearpage

\tableofcontents

\input{sec-introduction.tex}

\input{sec-beta-decays-theory_v2.tex}

\input{sec-observables}

\input{sec-methodology.tex}

\input{sec-results.tex}

\input{sec-summary}

\section*{Acknowledgements}

V.~P. acknowledges support from the INT and the NSF FRHTP programme under award No. PHY-2402275. 
E.~H. and J.~R. are supported by the Swiss National
Science Foundation.
L.~G. is supported by the Dutch Research Council (NWO) in form of an M1 grant. 
J.~d.~V. is funded by the Dutch Research Council (NWO) in the form of a VIDI grant and by the European Union (ERC, CRUNS, 101230525). 
Views and opinions expressed are, however,  those of the author(s) only and do not necessarily reflect those of the European Union or the European Research Council. Neither the European Union nor the granting authority can be held responsible for them.

\bibliographystyle{utphys}
\bibliography{BetaDecaysFCCee}

\end{document}

%% file: sec-introduction.tex
\section{Introduction}

The Standard Model (SM) of particle physics, despite its remarkable phenomenological 
success, leaves a number of fundamental questions unanswered, from the origin of neutrino 
masses to the nature of dark matter and the dynamics of electroweak symmetry breaking. 
In the absence of direct evidence for new resonances at the LHC, the indirect search for 
New Physics beyond the Standard Model (BSM) through precision measurements has become a 
central pillar of the high-energy physics programme. 

The Standard Model Effective Field  Theory (SMEFT)~\cite{Brivio:2017vri,Isidori:2023pyp,Grzadkowski:2010es} provides the natural framework for 
such a programme: under the assumption that BSM degrees of freedom are much heavier 
than the electroweak scale, their effects on SM observables can be systematically 
organised as a tower of higher-dimensional operators suppressed by powers of an 
ultraviolet (UV) scale $\Lambda$. 
The Wilson coefficients of these operators then encode the  imprint of New Physics in a model-independent fashion and can be constrained  simultaneously through global analyses combining a broad range of experimental inputs~\cite{Ethier:2021bye,Celada:2024mcf,terHoeve:2025omu,deBlas:2025xhe,Mantani:2025bqu,Ellis:2020unq,Almeida:2021asy,Bartocci:2023nvp,Elmer:2023wtr,ATLAS:2026fyh,Smolkovic:2026cba,Aebischer:2025qhh,CMS:2025ugn,Ethier:2021ydt,vanBeek:2019evb,Bissmann:2020mfi,Cirigliano:2016nyn}.

Multiple studies, {\it e.g.}~\cite{deBlas:2025gyz,Armadillo:2026mvp,Jung:2020uzh,Allwicher:2023shc,Maura:2024zxz,deBlas:2016nqo,DeBlas:2019qco,deBlas:2019rxi,deBlas:2021jlt,deBlas:2022ofj,Celada:2026uiw,Allwicher:2025mvd,Greljo:2025ggc,Altmann:2025feg,Cornet-Gomez:2025jot}, have quantified the  sensitivity of future high-energy colliders to dimension-six SMEFT operators, where legacy LEP and LHC measurements are combined with 
projections for the high-luminosity LHC (HL-LHC) and for the next generation of 
electron--positron facilities, including the Future Circular Collider  (FCC-ee)~\cite{FCC:2018evy,FCC:2025lpp,FCC:2025uan} prioritised by the Update of the European Strategy for Particle Physics~\cite{deBlas:2025gyz}.
The recent {\sc\small SMEFiT}~\cite{Giani:2023gfq,Hartland:2019bjb} study presented in~\cite{Armadillo:2026mvp} interprets FCC-ee projections  through global SMEFT fits including 
renormalisation-group evolution (RGE)~\cite{terHoeve:2025gey}, linear and quadratic EFT contributions, and NLO QCD and electroweak corrections to the EFT cross-sections whenever available. 
The resulting projections are further translated into  bounds on representative UV-complete extensions of the SM, demonstrating a comprehensive  picture of the New Physics reach of future collider precision programmes.

A complementary window into BSM Physics is provided by high-precision low-energy  observables such as semileptonic 
charged-current processes, including superallowed nuclear $\beta$-decays, neutron $\beta$-decay, mirror transitions, and pion and kaon $\beta$-decays~\cite{Cirigliano:2013xha,Gonzalez-Alonso:2018omy,Falkowski:2020pma}.
These  observables probe similar dimension-six SMEFT operators as those modifying charged currents at colliders, but through a different kinematic regime, with other hadronic and nuclear inputs, and with experimental uncertainties at the per-mille level. 
As such, they offer powerful constraints on a subset of SMEFT operators, in particular the 
two-quark--two-lepton and modified $W\ell\nu$ vertex operators, synergetic with those obtained from collider 
measurements~\cite{Falkowski:2017pss,Falkowski:2021bkq}. 

Recent interest in possible New Physics effects entering low-energy processes has been fuelled by the so-called ``Cabibbo Angle Anomaly'' (CAA), a $\sim$$3\sigma$ tension in the first-row  unitarity test of the CKM matrix~\cite{Hardy:2020qwl,Seng:2020wjq,Crivellin:2020lzu,Cirigliano:2022yyo} which is found from the best-fit values of $V_{ud}$ and $V_{us}$ extracted from these processes.
This anomaly has further sharpened the case for a consistent treatment of $\beta$- and meson-decays  within global SMEFT fits.
For instance, the CKM matrix element $V_{ud}$, which is the most precisely known element of the quark mixing matrix, is itself extracted under  the assumption of SM-only charged-current dynamics. In the presence of non-zero Wilson  coefficients, the determination of $V_{ud}$ from superallowed $0^{+}\to 0^{+}$ transitions, from neutron decay, and from mirror transitions becomes deeply intertwined with the BSM parameter  space, such that treating $V_{ud}$ as an external input in collider-based SMEFT fits is no  longer theoretically consistent~\cite{Falkowski:2017pss,Descotes-Genon:2018foz,Cirigliano:2023nol}. 
A simultaneous determination of $V_{ud}$ together with the relevant Wilson coefficients  is therefore required both to obtain reliable BSM bounds and to assess the robustness of 
unitarity tests of the CKM matrix in the possible presence of New Physics. 
Crucially, superallowed transitions by themselves cannot disentangle shifts in $V_{ud}$ from BSM contributions to left-handed charged currents. 

Motivated by these considerations, in this work we extend the {\sc\small SMEFiT}  framework with a dedicated low-energy likelihood that includes the most precise current  measurements of superallowed $0^{+}\to 0^{+}$ nuclear  transition lifetimes, together with the associated SM theory predictions and their hadronic 
and nuclear uncertainties. 
This low-energy likelihood is interfaced with the existing LEP, (HL-)LHC, and FCC-ee likelihoods available in {\sc\small SMEFiT} through RGE evolution and the matching to the Low-Energy Effective Field Theory (LEFT) below the electroweak scale~\cite{Jenkins:2017jig,Jenkins:2017dyc,Dekens:2019ept},  ensuring a consistent treatment of operator running and mixing across scales.
The CKM matrix element $V_{ud}$ is then promoted to an independent degree of freedom and 
determined simultaneously with the  Wilson coefficients.
We consider two commonly adopted flavour structures, namely $U(3)^5$ and $U(2)_{q_L}\otimes U(2)_{u_R}\otimes U(3)_{d_R}\otimes U(3)_{\ell}\otimes U(3)_e$, leaving the extension to more general structures to future work. 

By means of this \smefit-based analysis pipeline, we  derive simultaneous constraints on $V_{ud}$ and the Wilson coefficients for different input datasets: a global dataset consisting of low-energy, LEP and LHC observables; the same dataset extended with HL-LHC projections; and finally by adding the complete projections for FCC-ee observables from $\sqrt{s}=91$ GeV to $\sqrt{s}=365$ GeV.
We demonstrate that, while current data (and the upcoming HL-LHC measurements) are unable to fully break the New Physics degeneracies entering the determination of $V_{ud}$ from low-energy observables, the FCC-ee constraints enable a robust determination of $V_{ud}$ from $\beta$-decays with a clean separation from possible New Physics contributions, without any appreciable loss in precision even in optimistic scenarios for the precision of future theory calculations.

The structure of this paper is as follows. 
First, Sect.~\ref{sec:beta_decays_smeft} provides an overview of the EFT framework and flavour assumptions used in this work. 
Section~\ref{sec:observables} describes the relevant low-energy observables from superallowed $\beta$-decays, together with a concise summary of the collider observables following~\cite{Armadillo:2026mvp}.
In Sect.~\ref{sec:methodology} we present the SMEFT analysis methodology, and how \smefit\,has been extended to extract SM parameters together with EFT coefficients.
The main results of this work are given in Sect.~\ref{sec:results}, where we validate our SM determination of $V_{ud}$ with the PDG 2026 average~\cite{ParticleDataGroup:2026mpi}, present simultaneous extractions of $V_{ud}$ and EFT coefficients for different input datasets, and assess the stability of the results upon variations of the fit settings.
Finally, in Sect.~\ref{sec:summary} we conclude and discuss some possible avenues for future work.
The results presented in this work are made publicly available as part of the open-source {\sc\small SMEFiT} framework.

%% file: sec-beta-decays-theory_v2.tex
\section{EFT framework}
\label{sec:beta_decays_smeft}

In this section we summarise  the SMEFT framework adopted in this work, including the matching to the LEFT, the flavour structures, and the RGE implementation.
We focus on its implications for superallowed $\beta$-decays and for hadron and lepton collider observables, both to be discussed in Sect.~\ref{sec:observables}.

\paragraph{SMEFT and LEFT.}
Above the electroweak symmetry-breaking scale, the SM can be extended with a general basis of effective interactions which in the SMEFT~\cite{Isidori:2023pyp} are parametrised as
\begin{align}
\mathcal L_\text{SMEFT} = \sum_{d\ge 5} \mathcal L_\text{SMEFT}^{(d)} \,\quad\, {\rm with} \,\quad\,
    \mathcal L_\text{SMEFT}^{(d)} = \sum_i C_i^{(d)}\mathcal O_i^{(d)}\,,
    \label{eq:smeft_expansion}
\end{align}
where $C_i^{(d)}$ are the dimensionful Wilson coefficients (WCs) appearing in the Lagrangian at mass-dimension $d$. 
All BSM effects are thus incorporated into the WCs, with the New Physics scale $\Lambda$ appearing as $C_i^{(d)}\propto \Lambda^{4-d}$. 
In this work we focus on dimension-six operators, which are the most relevant for the interpretation of the low-energy and collider datasets considered here.
That is, in the expansion Eq.~(\ref{eq:smeft_expansion}) we only consider those with $d=6$.
We define the SMEFT operators in the Warsaw basis~\cite{Grzadkowski:2010es}, though some of the degrees of freedom (DoF) entering the fit are constructed as linear combinations thereof following~\cite{Armadillo:2026mvp}.

At energies below the $Z$-boson mass, the Higgs field and the heavy gauge bosons are integrated out, with the relevant EFT expansion being the low-energy effective field theory (LEFT), where
\begin{align}
\label{eq:left}
\mathcal L_\text{LEFT} = \sum_{d\ge 4} \mathcal L_\text{LEFT}^{(d)} \,\quad\, {\rm with} 
\,\quad\,
    \mathcal L^{(d)}_\text{LEFT} = \sum_iL_i^{(d)}\bar{\mathcal O}^{(d)}_i\,,
\end{align}
and where $L_i^{(d)}$ are the (also dimensionful) low-energy coefficients (LECs) and $\bar{\mathcal O}^{(d)}_i$ are the corresponding operators. 
Note that the LEC expansion of Eq.~(\ref{eq:left}) may also contain SM contributions, which are non-zero even when $C_i^{(d)}=0$ in Eq.~(\ref{eq:smeft_expansion}).
In this work we are interested in connecting collider observables with superallowed $\beta$-decays, with the latter requiring the evaluation of the LECs entering Eq.~(\ref{eq:left}) at $\mu \sim 2$~GeV.
These are then ultimately matched onto an EFT with hadrons as the effective degrees of freedom, namely Heavy Baryon Chiral Perturbation Theory~\cite{Jenkins:1990jv,Bernard:1995dp}.

The renormalisation group evolution  of the WCs~\cite{Alonso:2013hga} in the SMEFT from the New Physics scale $\mu_0=\Lambda$ to the scale $\mu$ at which the observables are evaluated is carried out at one-loop within \smefit\,through its interface to the software package \wilson~\cite{Aebischer:2018bkb} as described in~\cite{terHoeve:2025gey} (see~\cite{Mantani:2026fao,Haisch:2026eyq,Born:2026tgm} for recent two-loop results).
 At $\mu= m_Z$, the SMEFT WCs can be  matched to the LEFT LECs~\cite{Jenkins:2017jig}. 
The resulting LEFT LECs are then run down to $\mu\sim 2$~GeV according to the LEFT RGE equations~\cite{Jenkins:2017dyc}, for which we use the package \rgevolve~\cite{Smolkovic:2026cba}, which provides fast running and matching using precomputed evolution matrices.
We have verified that, for a common choice of settings, the operator running and mixing of \wilson and \rgevolve\,lead to identical results and hence can be used interchangeably.

\paragraph{Flavour assumptions.}
Our baseline SMEFT analysis adopts flavour universality, with the symmetry group being
\begin{align}
\label{eq:u35}
U(3)^5\equiv U(3)_{q_L}\times U(3)_{u_R}\times U(3)_{d_R}\times U(3)_\ell\times U(3)_e \, .
\end{align}
This flavour structure allows for a cleaner and simpler analysis of the interplay between low-energy and high-energy observables. 
Eq.~(\ref{eq:u35}) is the largest flavour symmetry of the SM Lagrangian in the limit of vanishing Yukawa couplings, and it acts on the three generations of each of the five fermion representations~\cite{DAmbrosio:2002vsn}. 
This symmetry is usually imposed in the SMEFT context  as an exact invariance of the dimension-six Lagrangian, that is, neglecting the Yukawa spurions that would break it.
In this way, the Wilson coefficients are diagonal and universal in
flavour space. 

With the $U(3)^5$ universal flavour assumptions of Eq.~(\ref{eq:u35}), chirality-flipping structures and right-handed charged currents, which require an insertion of a Yukawa coupling, are forbidden.
For this reason, among the LEFT operators which are potentially relevant to the description of $\beta$-decays, only the left-handed vector current survives, while the scalar, tensor, and right-handed operators have no counterpart.
This implies that the only New Physics effects entering the $\beta$-decay amplitudes arise from modifications of the left-handed couplings.
We recall however that right-handed couplings may play an important role in the resolution of the CAA \cite{Cirigliano:2022yyo, Cirigliano:2023nol} and therefore it would be interesting to consider more general flavour assumptions in future work. 

For the interpretation of collider observables we adopt the same universal flavour
assumption of Eq.~(\ref{eq:u35}).
This structure is different from the baseline one used in the {\sc\small SMEFiT}
analyses of~\cite{Celada:2024mcf,Armadillo:2026mvp}, under which the Warsaw basis
operators are taken to be invariant under the following flavour symmetry:
\begin{equation}
  \mathcal{G}_F \;=\; U(2)_{q_L}\otimes U(2)_{u_R}\otimes U(3)_{d_R}\otimes
  U(3)_{\ell}\otimes U(3)_e \,,
  \label{eq:smefit_flavour}
\end{equation}
where the first two generations of the left-handed quark doublets and of the
right-handed up-type quarks transform as doublets, while the third generation is a
singlet.
This flavour structure is motivated by the possible special role in the New Physics
played by the top quark and was originally proposed by the LHC top quark Working
Group~\cite{Aguilar-Saavedra:2018ksv}.

Singling out the third generation in the quark sector is what allows the top Yukawa
coupling to be retained while all other SM Yukawa couplings are neglected, since only
$y_t$ is then compatible with $\mathcal{G}_F$ and the remaining ones would act as
spurions breaking it.
This statement concerns the SM Yukawa couplings entering the
renormalisation group equations, and not the dimension-six operators that modify the
Yukawa interactions: the latter are retained in the operator fitting basis for the bottom,
charm, muon, and tau, namely $\mathcal{O}_{b\varphi}$, $\mathcal{O}_{c\varphi}$,
$\mathcal{O}_{\mu\varphi}$ and $\mathcal{O}_{\tau\varphi}$, so as to account for the
corresponding LHC and FCC-ee sensitivity to the associated Higgs decay channels. 
In the lepton sector, the $U(3)_\ell \times U(3)_e$ symmetry group enforces lepton-flavour conservation and vanishing lepton masses, so that the two-lepton and four-lepton operators entering the EWPOs and the $\beta$-decay likelihood are flavour universal. 

The $\mathcal{G}_F$ flavour assumption in Eq.~(\ref{eq:smefit_flavour}) is less restrictive than the $U(3)^5$ symmetry, yet  nevertheless the two coincide for the operators relevant to $\beta$-decays.
Indeed, since $\beta$-decays involve only first-generation quarks and light leptons, they belong to the light-generation, flavour-universal subset
common to both flavour symmetry groups. 
In particular, we note that the SMEFT degrees of freedom (to be listed below in Table~\ref{tab:SMEFiT_and_U(3)^5}) and those required to describe low-energy observables map identically onto the corresponding \smefit\,coefficients, ensuring a consistent combination of the low-energy and collider likelihoods. 
For the latter, the additional flavour constraints of Eq.~(\ref{eq:u35}) are imposed at the runcard level.
In Sect.~\ref{subsec:stability} we assess the stability of our results with respect to the choice of flavour assumptions, by comparing results based on $U(3)^5$ with those based on $\mathcal{G}_F$.

Finally, we note that in the \smefit theory database, collider observables are parametrised in terms of the more general flavour symmetry group
\be
\label{eq:gfprime}
 \mathcal{G}_F^\prime =  
U(2)_{q_L}\otimes U(2)_{u_R}\otimes U(3)_{d_R}\otimes (U(1)_{\ell}\otimes U(1)_e)^3 \, ,
\ee
with the flavour universality in the leptonic sector leading to Eq.~(\ref{eq:smefit_flavour}) always imposed at the runcard level. 

\paragraph{EFT description of $\beta$-decays.}
Under the $U(3)^5$ universal flavour assumptions that we use in this work, the LEFT Lagrangian Eq.~(\ref{eq:left}) relevant to describe $\beta$-decays contains only left-handed currents, namely
\begin{align}
    \mathcal L_{\rm LEFT} \supset L^{V,LL(1111)}_{\nu edu}\left(\bar u \gamma^\mu d_L\right)\left(\bar e \gamma_\mu\nu_L\right) + \text{h.c.}\,,
    \label{eq:left_interaction}
\end{align}
hence depending on a single four-fermion (two-quark-two-lepton) operator, and
where the corresponding LEC $L^{V,LL(1111)}_{\nu edu}$ contains both SM and BSM contributions.
Following~\cite{Cirigliano:2023nol}, we rotate the left-handed up-type quarks from the interaction basis to the mass basis with the CKM matrix $V$,
\begin{align}
u_L^\text{(int)}\rightarrow V^\dagger u_L^\text{(mass)} \, ,
\end{align}
such that the relevant LEFT Lagrangian for a general flavour structure $prst$ transforms as
\begin{align}
    \mathcal L_{\rm LEFT} \supset &\,L^{V,LL(prst)}_{\nu edu}\left(\bar u_t \gamma^\mu P_L d_s\right)\left(\bar e_p \gamma_\mu P_L \nu_r\right)\nonumber\\
    =&\,L^{V,LL(prst)}_{\nu edu}V_{xt}\left(\bar u_x \gamma^\mu P_L d_s\right)\left(\bar e_p \gamma_\mu P_L \nu_r\right)\,,
\label{eq:mass_basis_rotation}
\end{align}
where $x$ stands for a mass eigenstate label. 
Hence, low-energy observables defined in the mass basis contain a combination of the LEFT LECs and CKM elements dictated by Eq.~(\ref{eq:mass_basis_rotation}).

At energies of $\mu\sim m_Z$, the LEFT Lagrangian Eq.~(\ref{eq:left_interaction}) is matched to the SMEFT. 
The corresponding matching relations involve the dimension-six SMEFT operators listed in Table~\ref{tab:operator_definitions}, see also Tables~\ref{tab:coefficient_rotation} and~\ref{tab:SMEFiT_and_U(3)^5} for additional details.
Starting from the general Warsaw basis, the  $U(3)^5$ flavour symmetry imposes a number of relations between the operators that contribute to $\beta$-decays,
\begin{align}
    C_{\varphi \ell}^{3(pr)}\propto \left(\mathbf{1}\right)_{pr}\,,\quad C_{\varphi q}^{3(pr)}\propto \left(\mathbf{1}\right)_{pr}\,,\quad C_{\ell q}^{3(prst)}\propto \left(\mathbf{1}_\ell\right)_{pr}\times \left(\mathbf{1}_q\right)_{st}\,,\quad C_{\ell\ell}^{(1221)} = C_{\ell\ell}^{(1331)} = C_{\ell\ell}^{(2332)}\,,
\end{align}
such that the WCs corresponding to fermion flavour-mixing operators vanish, while the flavour-diagonal WCs will have the same value, set by the dimensionful constant of proportionality, across flavours. 

The resulting matching relation between the LEFT LEC and the SMEFT WC relevant for $\beta$-decays, for the $\mathcal{G}_F^\prime$ flavour structure Eq.~(\ref{eq:gfprime}), is then given by
\begin{align}
    L_{\nu e du}^{V,LL(1111)} =& -\frac{2}{v_T^2} + 2\left(C^{3(1111)}_{\ell q} - C^{3(11)}_{\varphi \ell} - C^{3(11)}_{\varphi q}\right)\nonumber\\
    =&-2\sqrt{2}\GFexp - \left(C^{(2112)}_{\ell\ell} + C^{(1221)}_{\ell\ell} - 2C^{3(22)}_{\varphi \ell} - 2C^{3(11)}_{\varphi \ell}\right)+ 2\left(C^{3(1111)}_{\ell q} - C^{3(11)}_{\varphi \ell} - C^{3(11)}_{\varphi q}\right)\nonumber\\
    =&-2\sqrt{2}\GFexp - \left(C^{(2112)}_{\ell\ell} + C^{(1221)}_{\ell\ell} - 2C^{3(22)}_{\varphi \ell}  -2C^{3(1111)}_{\ell q} + 2C^{3(11)}_{\varphi q}\right)\nonumber\\
    \xrightarrow[]{\text{\smefit~\text{DOFs}}}& \underbrace{-2\sqrt{2}\GFexp}_{L_0} + \underbrace{2\left(c_{q\ell_1}^{(3)} - c_{\varphi q}^{(3)}-c'_{\ell\ell} + c_{\varphi \ell_2}^{(3)}\right)}_{L'^{(1111)}}\,,
    \label{eq:LVLL}
\end{align}
where in the final row the SMEFT WCs are expressed in terms of the \smefit\,degrees of freedom listed in Tables~\ref{tab:operator_definitions} and~\ref{tab:SMEFiT_and_U(3)^5}. 
As mentioned above, the LEC can be separated into a SM contribution $L_0$,
\be
L_0 \equiv -2\sqrt{2}\GFexp \, ,
\label{eq:L0}
\ee
and a BSM contribution denoted by
\begin{align}
\label{eq:smeft_combination_lprime}
L'^{(1111)}\equiv 2\left(c_{q\ell_1}^{(3)} - c_{\varphi q}^{(3)}-c'_{\ell\ell} + c_{\varphi \ell_2}^{(3)}\right) \xrightarrow[\text{}]{U(3)^5}2\left(c_{q\ell}^{(3)} - c_{\varphi q}^{(3)}-c'_{\ell\ell} + c_{\varphi \ell}^{(3)}\right)\, ,
\end{align}
where the RHS corresponds to the matching relation under the $U(3)^5$ flavour symmetry.
The same result is obtained under the $\mathcal{G}_F$ group.
Therefore, from the point of view of the interpretation of $\beta$-decays, the relevant EFT coefficients in our baseline $U(3)^5$ flavour structure are
\begin{align}
\label{eq:operators_smeft_matching_mz}
    c_{q\ell}^{(3)}, \, c_{\varphi \ell}^{(3)}, \, c_{\varphi q}^{(3)}, \, c_{\ell \ell}^\prime \, ,
\end{align}
which are also constrained at lepton and hadron colliders~\cite{Armadillo:2026mvp}.

\begin{table}[t] 
  \begin{center}
    \renewcommand{\arraystretch}{1.8}
    \begin{tabularx}{\textwidth}{Xlllcc}
      \toprule
      $C_i$ (Warsaw) $\quad$ & $C_i$ (SMEFiT) $\quad$ & $\mathcal{O}_i$ $\quad$ & Definition & DoF ($\mathcal{G}_F^\prime$)
      & DoF ($\mathcal{G}_F$)\\
      \midrule
      $C^{3(iijj)}_{\ell q}$ & $c_{q\ell_j}^{(3)}$ & $\Op{q \ell_j}^{(3)}$ &
      $\sum\limits_{\sss i=1,2}
      \left(\bar{\ell}_j \gamma_\mu \tau_{\sss I} \ell_j\right)
      \left(\bar{q}_i \gamma^\mu \tau^{\sss I} q_i\right)$ & 3 & 1
      \\

      $C^{3(33jj)}_{\ell q}$ & $c_{Q\ell_j}^{(3)}$ & $\Op{Q \ell_j}^{(3)}$ &
      $\left(\bar{\ell}_j \gamma_\mu \tau_{\sss I} \ell_j\right)
      \left(\bar{Q} \gamma^\mu \tau^{\sss I} Q\right)$ & 3 & 1
      \\

      $C^{3(ii)}_{\varphi q}$ & $c_{\varphi q}^{(3)}$ & $\Op{\varphi q}^{(3)}$ &
      $\sum\limits_{\sss i=1,2}
      i\big(\varphi^\dagger\,\lra{D}_\mu\,\tau_{\sss I}\varphi\big)
      \big(\bar{q}_i\,\gamma^\mu\,\tau^{\sss I}q_i\big)$ & 1 & 1
      \\

      $C^{3(33)}_{\varphi q}$ & $c_{\varphi Q}^{(3)}$ & $\Op{\varphi Q}^{(3)}$ & 
      $i\big(\varphi^\dagger\,\lra{D}_\mu\,\tau_{\sss I}\varphi\big)
      \big(\bar{Q}\,\gamma^\mu\,\tau^{\sss I}Q\big)$ & 1 & 1
      \\

      $C^{3(jj)}_{\varphi \ell}$ & $c_{\varphi \ell_j}^{(3)}$ &
      $\Op{\varphi \ell_j}^{(3)}$ &
      $i\big(\varphi^\dagger\lra{D}_\mu\,\tau_{\sss I}\varphi\big)
      \big(\bar{\ell}_j\,\gamma^\mu\,\tau^{\sss I}\ell_j\big)$ & 3 & 1
      \\

      $C^{(kjjk)}_{\ell \ell}$ & $c_{\ell \ell}^{\prime}$ &
      $\qq{}{\ell \ell}{kjjk}$ &
      $\left(\bar \ell_k\gamma_\mu \ell_j\right)
      \left(\bar \ell_j\gamma^\mu \ell_k\right)$ & 6  & 1
      \\
      \midrule
      $C^{1(iijj)}_{\ell q}$ & $c_{q\ell_j}^{(1)}$ & $\Op{q \ell_j}^{(1)}$ &
      $\sum\limits_{\sss i=1,2}
      \left(\bar{\ell}_j\gamma_\mu\ell_j\right)
      \left(\bar{q}_i\gamma^\mu q_i\right)$ & - & -
      \\
      
      $C^{1(33jj)}_{\ell q}$ & $c_{Q\ell_j}^{(1)}$ & $\Op{Q \ell_j}^{(1)}$ &
      $\left(\bar{\ell}_j\gamma_\mu\ell_j\right)
      \left(\bar{Q}\gamma^\mu Q\right)$ & -  & -
      \\

      $C^{1(ii)}_{\varphi q}$ & $c_{\varphi q}^{(1)}$ & $\Op{\varphi q}^{(1)}$ &
      $\sum\limits_{\sss i=1,2}
      i\big(\varphi^\dagger\,\lra{D}_\mu\,\varphi\big)
      \big(\bar{q}_i\,\gamma^\mu q_i\big)$ & - & -
      \\

      $C^{1(33)}_{\varphi q}$ & $c_{\varphi Q}^{(1)}$ & $\Op{\varphi Q}^{(1)}$ &
      $i\big(\varphi^\dagger\,\lra{D}_\mu\,\varphi\big)
      \big(\bar{Q}\,\gamma^\mu Q\big)$ & - & -
      \\
      \bottomrule
    \end{tabularx}
    \vspace{0.2cm}
    \caption{The definition of the SMEFT operators considered in this analysis and relevant to the description of $\beta$-decays, with the corresponding Wilson coefficients expressed both in the Warsaw basis and in the \smefit fitting basis notation. 
    In each case, we indicate the number of degrees of freedom assuming as flavour symmetry either
    $\mathcal{G}_F^\prime$ 
    or $\mathcal{G}_F$ (while in the case of $U(3)^5$ one further equates $q$ with $Q$).
    The number of DoF reduces to 6 with the $\mathcal{G}_F$ flavour assumption and then to 4 under $U(3)^5$, which treats equally all quark generations, see also Table~\ref{tab:SMEFiT_and_U(3)^5}.
    Note that the SU(2)$_L$ singlet operators (bottom four rows) are set to zero in both cases, see Table~\ref{tab:coefficient_rotation}, since they are not relevant for the matching with the LEFT coefficient  $L'^{(1111)}$.
   These singlet operators are listed here for completeness since they are part of the DoF definition in the \smefit fitting basis.}
  \label{tab:operator_definitions}
  \end{center}
\end{table}

\begin{table}[t]
  \begin{center}
    \renewcommand{\arraystretch}{1.8}
    \begin{tabularx}{0.7\textwidth}{Xl}
      \toprule
      Coefficient (\smefit) & Definition \\
      \midrule
      $c_{\varphi q}^{(-)}$
      & $c_{\varphi q}^{(1)} - c_{\varphi q}^{(3)}$
      \\
      $c_{\varphi Q}^{(-)}$
      & $c_{\varphi Q}^{(1)} - c_{\varphi Q}^{(3)}$
      \\
      $c_{q\ell_j}^{(-)}$
      & $c_{q\ell_j}^{(1)} - c_{q\ell_j}^{(3)}$
      \\
      $c_{Q\ell_j}^{(-)}$
      & $c_{Q\ell_j}^{(1)} - c_{Q\ell_j}^{(3)}$
      \\
      \bottomrule
    \end{tabularx}
    \vspace{0.4cm}
    \caption{Definition of the DoFs entering the \smefit fitting basis as linear combinations of the SU$(2)_L$ singlet and triplet Higgs-difermion operators, see Table~\ref{tab:operator_definitions} for their corresponding definitions.
    For the fits presented in this work, we set the SU$(2)_L$ singlet coefficients to zero, that is, we impose $c_{\varphi q}^{(1)}=c_{\varphi Q}^{(1)}=c_{q\ell_j}^{(1)}=c_{Q\ell_j}^{(1)}=0$ hence leading to the relations displayed in the last column of Table~\ref{tab:SMEFiT_and_U(3)^5}.
    }
    \label{tab:coefficient_rotation}
  \end{center}
\end{table}

\begin{table}[t] 
  \begin{center}
    \renewcommand{\arraystretch}{1.8}
        \begin{tabularx}{\textwidth}{lllX}
          \toprule
           \multicolumn{4}{c}{LEFT matching} \\
           \midrule
        $\qquad$ $\qquad$ $\qquad$ $\qquad$&  $\mathcal{G}_F^\prime$ & $\qquad$ $\qquad$ $\qquad$$U(3)^5$ $\qquad$$\qquad$ & $\mathcal{G}_F$ + SMEFiT DoFs \\
        \midrule
        \midrule
        $\quad$ & $c_{q\ell_1}^{(3)}$ & $\qquad$ $\qquad$ $\qquad$$c_{q\ell}^{(3)}$ &
        $c_{q\ell}^{(3)} = -c_{q\ell}^{(-)}, \,c_{Q\ell}^{(3)} = - c_{Q\ell}^{(-)}$,   
        \\
        $\quad$ & $c_{\varphi \ell_2}^{(3)}$ & $\qquad$ $\qquad$ $\qquad$ $c_{\varphi \ell}^{(3)}$ &
        $c_{\varphi \ell}^{(3)}$
        \\
        $\quad$ & $c_{\varphi q}^{(3)}$ & $\qquad$ $\qquad$ $\qquad$ $c_{\varphi q}^{(3)}$ &
        $c_{\varphi q}^{(3)}=- c_{\varphi q}^{(-)}, \, c_{\varphi Q}^{(3)}=- c_{\varphi Q}^{(-)}$
        \\
        $\quad$ & $c_{\ell \ell}^{1221}$ & $\qquad$ $\qquad$ $\qquad$ $c_{\ell \ell}^{\prime}$ &
        $c_{\ell \ell}^{\prime}$
        \\       
    \midrule
        DoF in fits & - & $\qquad$ $\qquad$ $\qquad$  4 & 6
        \\       
       \bottomrule
        \end{tabularx}
        \vspace{0.2cm}
        \caption{The SMEFT DoFs relevant for the matching with the LEFT LEC describing $\beta$-decays, Eq.~(\ref{eq:LVLL}), for the different flavour structures considered in this work.
        In the fourth column we indicate the relations between \smefit DoFs which are imposed at the fit level, see also Table~\ref{tab:coefficient_rotation}.
        The last row lists the active DoF for each configuration.
        }
\label{tab:SMEFiT_and_U(3)^5}
\end{center}
\end{table}

In deriving Eq.~(\ref{eq:LVLL}), we have used that the SM contribution in Eq.~(\ref{eq:L0})
is proportional to the experimentally measured value of the Fermi constant extracted from muon decays~\cite{Eberhart:2026klz},
\be
\GFexp = 1.166 378 59(59)\times 10^{-5}\text{ GeV}^{-2}\,.
\ee 
This experimental measurement $\GFexp$ contains potential BSM contributions through the Wilson coefficients $C^{3(11)}_{\varphi\ell},\,C^{3(22)}_{\varphi\ell}$, and $C^{(2112)}_{\ell\ell} \left(=C^{(1221)}_{\ell\ell}\right)$~\cite{Jenkins:2017jig,Cirigliano:2023nol} as follows: 
\begin{align}
    \frac{4\GFexp}{\sqrt{2}} = \frac{2}{v_T^2} - C^{(2112)}_{\ell\ell} - C^{(1221)}_{\ell\ell} + 2C^{3(22)}_{\varphi\ell} + 2C^{3(11)}_{\varphi\ell}\,,
    \label{eq:GFshift}
\end{align}
where we have defined 
\begin{align}
v_T\equiv\left(1 + \frac{3C_H\,v^2}{8\lambda}\right)v \, ,
\end{align}
as the SMEFT-modified Higgs vacuum expectation value, with $v=246.22$ GeV being the SM Higgs vev.
However, since the Higgs vev is also extracted from measurements and cannot be directly measured, we will equate $G_F = \left(\sqrt 2 v_T^2\right)^{-1}$ and refer to this as the SM value of Fermi's constant.
That is, the relation between the SM Fermi constant and its experimentally measured value is
\begin{align}
 \frac{4\GFexp}{\sqrt{2}} =  \frac{4G_F}{\sqrt{2}} - \lp C^{(2112)}_{\ell\ell} + C^{(1221)}_{\ell\ell} - 2C^{3(22)}_{\varphi\ell} - 2C^{3(11)}_{\varphi\ell} \rp \,,
    \label{eq:GFshift_v2}
\end{align}
which enables expressing the relevant LEC $ L_{\nu e du}^{V,LL(1111)}$ in terms of the experimentally measured value $\GFexp$ which defines its SM contribution $L_0$.

Combining these results and rotating to the mass (physical) basis, we find that superallowed $\beta$-decays are sensitive to an effective CKM parameter given by
\be
\label{eq:VuDeff}
    \widetilde V_{ud} = \,V_{ud}\left(1 - \frac{\sum_i V_{ui}L'^{(1 1 1 i)}}{2\sqrt{2}G_F^\text{(exp)}V_{ud}}\right)
    \xrightarrow[]{U(3)^5}\,V_{ud}\left(1 - \frac{L'^{(1111)}}{2\sqrt{2} \GFexp}\right) \, ,
\ee
where the $i=\{1,2,3\}$ elements correspond to $\{V_{ud},V_{us},V_{ub}\}$ CKM elements. Under the $U(3)^5$ universal flavour assumptions, the second term vanishes unless $i=1$.
We then find that superallowed $\beta$-decays, to be described in more detail in Sect.~\ref{sec:observables}, depend on both $V_{ud}$ and on the SMEFT combination in Eq.~(\ref{eq:smeft_combination_lprime}).
From Eq.~(\ref{eq:VuDeff}) it becomes clear that by using only low-energy data it is not possible to disentangle shifts in $V_{ud}$ from BSM effects arising from a non-zero value of $L'^{(1111)}$~\cite{Falkowski:2020pma}.

\paragraph{DoFs in the fits.}
We conclude this section with a brief summary of the DoFs entering the fits presented in this work.
Table~\ref{tab:operator_definitions} collects the definitions of the SMEFT operators
relevant for the interpretation of superallowed $\beta$-decays, both in the Warsaw and in
the \smefit\,fitting bases, while Table~\ref{tab:SMEFiT_and_U(3)^5} summarises the
resulting DoFs for each of the flavour structures considered in this work.
Under the general $\mathcal{G}_F^\prime$ assumption, the matching of Eq.~(\ref{eq:LVLL})
singles out four specific flavour components, namely $c^{(3)}_{q\ell_1}$,
$c^{(3)}_{\varphi \ell_2}$, $c^{(3)}_{\varphi q}$, and $c^{1221}_{\ell\ell}$.
Imposing the universal $U(3)^5$ symmetry of Eq.~(\ref{eq:u35}) identifies all lepton and
quark generations, and these become the four independent coefficients of
Eq.~(\ref{eq:operators_smeft_matching_mz}) which define our baseline fits.
Under the less restrictive $\mathcal{G}_F$ symmetry of Eq.~(\ref{eq:smefit_flavour}),
instead, the third quark generation is no longer identified with the first two, and the
number of active DoFs increases to six, with $c^{(3)}_{Q\ell}$ and
$c^{(3)}_{\varphi Q}$ becoming independent.
These two additional coefficients do not enter the $\beta$-decay matching, since
superallowed transitions involve only first-generation quarks, but they do affect the
collider observables, in particular those sensitive to the top-quark sector.
Comparing the two flavour structures therefore provides a validation of the
stability of our results, which we present in Sect.~\ref{subsec:stability}.

In both cases (fits with $U(3)^5$ and with $\mathcal{G}_F$) we set the $SU(2)_L$ singlet coefficients to zero, see
Table~\ref{tab:coefficient_rotation}, which within the \smefit\,fitting basis translates
into the relations displayed in the last column of Table~\ref{tab:SMEFiT_and_U(3)^5}. 
The rationale is that the singlet operators generate only neutral-current structures,
and hence contribute at tree level neither to the charged-current LEFT coefficient
$L'^{(1111)}$ of Eq.~(\ref{eq:smeft_combination_lprime}) nor to the extraction of
$\GFexp$ from muon decay, Eq.~(\ref{eq:GFshift}).
The same reasoning applies to the four-lepton invariant $c_{\ell\ell}$, which is likewise set to zero, since only the crossed combination $c'_{\ell\ell}$ enters Eq.~(\ref{eq:GFshift}).
These conditions are imposed at the fit scale $\mu_0=10$~TeV, so that any singlet component generated by the RGE at lower scales is fully determined by the operators listed above, which therefore form a closed system connecting collider and low-energy
observables.
This is a choice of UV scenario, rather than a consequence of the flavour symmetry.
Indeed, activating the singlet directions would enlarge the collider parameter space
without adding sensitivity to $\beta$-decays, and we leave such an extension of our analysis to future work as outlined in Sect.~\ref{sec:summary}.

%% file: sec-observables.tex
\section{Low- and high-energy observables}
\label{sec:observables}

In this work we combine low-energy observables, in particular superallowed $\beta$-decays, with high-energy processes which function as complementary probes of BSM physics.
For collider observables, we use both all available world data (LEP and LHC) as well as projections for the HL-LHC and the FCC-ee.
We note that the present analysis of superallowed $\beta$-decays and their connection with LHC observables is heavily based on~\cite{Bhattacharya:2011qm,Cirigliano:2012ab,Cirigliano:2013xha,Cirigliano:2023nol}.
We describe the various considered datasets in turn.

\paragraph{Superallowed $\beta$-decay rates.}
Superallowed $\beta$-decay transitions play a central role in particle physics. 
In superallowed $\beta$-decays, a nucleus with zero total angular momentum and positive parity, namely a $0^+$ state, decays into another nucleus with the same quantum numbers. 
While the weak interaction has a V$-$A structure, only the vector part contributes to superallowed transitions. 
This greatly simplifies the theoretical description because, in the isospin limit, the vector current is not renormalized by QCD and the associated nuclear matrix element is fixed by group theory and is therefore nucleus-independent. 

The lifetime of superallowed $\beta$-decays depends sensitively on the up-down CKM element $V_{ud}$.
The current global average from the 2026 PDG~\cite{ParticleDataGroup:2026mpi} is given by 
\begin{align}
\label{eq:vud}
V_{ud}=0.97367 \pm 0.00032~(\pm\, 0.033\%)\, ,
\end{align}
and is obtained from the average over a large number of different isotopes. 
$V_{ud}$ can also be measured in pion and neutron decays, but there experimental uncertainties are larger than for superallowed transitions. 
While the current precision in the extraction of $V_{ud}$, Eq.~(\ref{eq:vud}), is impressive, the measurements of superallowed decay rates are even more precise.
However, at the current level of precision, the interpretation of these measurements is limited by the accuracy of radiative corrections (RC) entering theory calculations. 
These RC arise from electromagnetic interactions between the outgoing electron and the final-state nucleus, and from electromagnetic and isospin-breaking interactions within the atomic nuclei. 

Superallowed transitions involve only the vector part of the weak interaction and the resulting nuclear matrix elements are trivial up to effects from RC. 
The discussion of RC in $\beta$-decays has a long history, starting from Fermi~\cite{Fermi:1934hr}, and is periodically reviewed by Towner and Hardy (see also \cite{Seng:2026tgw} for a recent review). 
Their most recent compilation of RC is given in~\cite{Hardy:2020qwl}, and we  discuss new developments and their impact below. 
The strategy of Towner and Hardy is to separate nucleus-dependent quantities (theoretical and experimental) from nucleus-independent quantities through the master formula 
\begin{align}
\label{eq:Ft_i}
    \mathcal{F}t = \frac{2\pi^3\ln(2)}{ m_e^5}C_V^{-2}\,,
\end{align}
where $\mathcal{F} t$ is a combination of phase-space corrections and partial half-lives discussed below, and the (isotope universal) vector coupling $C_V$ is given by 
\begin{align}
\label{eq:CV_BSM}
    C_V &= \sqrt2\GFexp \widetilde V_{ud}\,\sqrt{1 + \Delta_R^V}\,. 
\end{align}
The RHS of Eq.~\eqref{eq:Ft_i} depends on the experimentally measured value of the Fermi constant, $G_F^{(\rm exp)}$, the BSM-dressed CKM element, $\widetilde V_{ud}$, defined in Eq.~(\ref{eq:VuDeff}), and the (universal) radiative correction, $\Delta_R^V$.
From Eq.~(\ref{eq:CV_BSM}) we observe that since the lifetime of all $\beta$-decays is proportional to the effective CKM element $\widetilde V_{ud}$ it is impossible to disentangle genuine shifts in the value of $V_{ud}$ from possible BSM effects by means of $\beta$-decay lifetime measurements alone. 
The isotope-independent $\Delta_R^V$ describes the universal nucleus-independent RC that affects all $\beta$-decays in the same way.
Its precise value has been under intense scrutiny in recent years~\cite{Seng:2018yzq, Czarnecki:2019mwq,Cirigliano:2023fnz}, and in this work we use the value
\begin{align}
\Delta_R^V= 0.02471 \pm 0.00025~(\pm\, 1.0\%)\, ,
\end{align}
as considered in~\cite{Cirigliano:2023fnz}. 

Since the RHS of Eq.~(\ref{eq:Ft_i}) is nucleus independent, the LHS should be as well. 
While the nuclear matrix elements are the same for different isotopes, the decay rates also involve phase space factors determined by the masses of the initial- and final-state nuclei (hence with different $Q$-values). 
They also depend on the charge of the final-state nucleus as the electron wave functions are distorted by the Coulomb field of the nucleus. 
This isotope-dependent phase space factor is called $f_i$, where $i$ labels the isotope.  
Once we multiply the phase space factor $f_i$ by the observed half life $t_i$ for the corresponding isotope, one obtains almost isotope independent $f_i t_i$ values, with deviations from exact independence at the percent level. 

These deviations are caused by additional RC not captured already by the universal term $\Delta_R^V$ in Eq.~(\ref{eq:CV_BSM}).
To account for these, Towner and Hardy introduce the following corrections
\begin{equation}
\mathcal F t = \left(1+\delta_R'\right)\left(1+\delta_{\rm NS} - \delta_C\right)f_i\,t_i\,,
\label{eq:isotope-dep-corrections}
\end{equation}
where the various $\delta$'s indicate transition-specific RCs. 
The $\delta'_R$ computed at leading order in $\mathcal O(\alpha_{\mathrm{em}})$ is given by the universal Sirlin function~\cite{Sirlin:1967zza} but at higher order depends on the final state atomic number $Z$. 
$\delta_C$ arises from isospin-breaking interactions in the nuclear wave functions. 
The third isotope-dependent RC is denoted by  $\delta_{\rm NS}$  and arises from  nuclear-structure effects, for example, from photon exchange between the electron and a proton in the nucleus that was not involved in the weak transition. 
$\delta_{\rm NS}$ is typically split into an electron-energy independent part, $\delta^{(0)}_{\rm NS}$, and an electron-energy dependent part, $\delta_{\rm NS, E}$. 

The various RCs entering Eq.~(\ref{eq:isotope-dep-corrections}) have been evaluated for a large range of transitions, finding that as expected the resulting $\mathcal F t$ values are isotope-independent \cite{Hardy:2020qwl}. 
This result then allows for a direct extraction of $\widetilde V_{ud}$ from Eq.~\eqref{eq:Ft_i} by combining data from the 15 most precise superallowed transitions. We collect the data used in the present analysis in Table~\ref{tab:beta_decay}. 
For each isotope, we indicate the measured phase-space corrected lifetime $\mathcal{F}t$ (labelled $\mathcal F t^{\rm (HT)}$) and uncertainty, the corresponding $Q$-value, and the value of the theory uncertainty $\Delta \delta_R'$ associated to the calculation of $\delta'_R$.

The calculation of these isotope-specific RCs induces theoretical uncertainties that are correlated over the different transitions. 
For the inclusion of superallowed $\beta$-decays in our combined likelihood we follow the procedure outlined in~\cite{Falkowski:2020pma} which was also used in earlier SMEFT analyses of low-energy data~\cite{Cirigliano:2023nol}. 
In this approach, the effects of the isotope-dependent RC uncertainties are taken into account in the likelihood by means of univariate Gaussian nuisance parameters $\eta_j$.
That is, the quantities which are compared to theoretical predictions, the phase-space corrected lifetimes  $\mathcal F t$ of Eq.~(\ref{eq:Ft_i}), are the extracted $\mathcal F t^{\rm (HT)}$ values from  \cite{Towner:2007np} (see Table~\ref{tab:beta_decay}) accounting for three theory nuisance parameters, namely
\begin{align}
\label{eq:ftpred_vs_ftexp}
    \mathcal F t = \mathcal F t^{\rm(HT)}\left(1 + \eta_1\Delta\delta'^i_R + \eta_2\Delta\delta_{\rm NS}^{(0)} +\eta_3\Delta\delta_{\rm NS,E}^i\right)\,,
\end{align}
where the parameters $\Delta\delta_X$ represent theoretical uncertainties entering the isotope-specific RC of Eq.~(\ref{eq:isotope-dep-corrections}) and estimated in the literature following~\cite{Hardy:2020qwl, Gorchtein:2018fxl, Seng:2018qru}. 
In this work we use $\Delta\delta_{\rm NS,E}^i=0.8\cdot10^{-4}$~ $Q_i$/MeV as a function of the $Q$-value (see Table~\ref{tab:beta_decay}) of the corresponding decay~\cite{Gorchtein:2018fxl}. 
We then set 
\be
\Delta\delta_{\rm NS}^{(0)} =3.3\cdot10^{-4}
\label{eq:delta_NS_1}
\ee
as an isotope-independent uncertainty~\cite{Seng:2018qru,Hardy:2020qwl}.
We note that this assumption used in~\cite{Falkowski:2020pma,Cirigliano:2023nol} is somewhat optimistic and we discuss in Sect.~\ref{sec:results} how our results change if a more conservative choice is adopted. 
Finally, we follow \cite{Falkowski:2020pma} and set $\Delta\delta'^i_R$ to be equal to one third of the $Z^2\alpha^3$ term in $\delta'^i_R$ ~\cite{Towner:2007np, Gonzalez-Alonso:2018omy,Hardy:2014qxa}.
This resulting uncertainty is given in the last column of Table~\ref{tab:beta_decay}.
We do not include a dedicated nuisance parameter associated to $\delta_C$ in Eq.~(\ref{eq:isotope-dep-corrections}) since this uncertainty is considered to be  subdominant~\cite{Hardy:2020qwl}. 

\begin{table}[t!]
\centering
\renewcommand{\arraystretch}{1.5}
\begin{tabular}{rcccc}
\toprule
Isotope & $\mathcal F t^{\rm (HT)}$ (s) & $\sigma_{\mathcal Ft}$ (s) & $Q$ (MeV) & $\Delta\delta'_R$ \cite[table V]{Towner:2007np} \\
\midrule
\midrule
$^{10}$C & 3075.7 & 4.4 & 1.908 & $9.0\times 10^{-5}$ \\
$^{14}$O & 3070.2 & 1.9 & 2.831 & $7.7\times 10^{-5}$ \\
$^{22}$Mg & 3076.2 & 7.0 & 4.125 & $6.7\times 10^{-5}$ \\
$^{26m}$Al & 3072.4 & 1.1 & 4.233 & $6.7\times 10^{-5}$ \\
$^{26}$Si & 3075.4 & 5.7 & 4.841 & $6.3\times 10^{-5}$ \\
$^{34}$Cl & 3071.6 & 1.8 & 5.492 & $6.0\times 10^{-5}$ \\
$^{34}$Ar & 3075.1 & 3.1 & 6.062 & $5.7\times 10^{-5}$ \\
$^{38m}$K & 3072.9 & 2.0 & 6.044 & $5.7\times 10^{-5}$ \\
$^{38}$Ca & 3077.8 & 6.2 & 6.612 & $5.7\times 10^{-5}$ \\
$^{42}$Sc & 3071.7 & 2.0 & 6.426 & $5.7\times 10^{-5}$ \\
$^{46}$V & 3074.3 & 2.0 & 7.052 & $5.3\times 10^{-5}$ \\
$^{50}$Mn & 3071.1 & 1.6 & 7.634 & $5.0\times 10^{-5}$ \\
$^{54}$Co & 3070.4 & 2.5 & 8.244 & $5.0\times 10^{-5}$ \\
$^{62}$Ga & 3072.4 & 6.7 & 9.181 & $4.7\times 10^{-5}$ \\
$^{74}$Rb & 3077.0 & 11.0 & 10.417 & $4.3\times 10^{-5}$ \\
\bottomrule
\end{tabular}
\vspace{0.2cm}
\caption{The experimental measurements of superallowed $\beta$-decays considered in this analysis and taken from~\cite{Hardy:2020qwl,Towner:2007np}.
For each isotope, we indicate the measured phase-space corrected lifetime $\mathcal{F}t^{\rm(HT)}$ and uncertainty, the corresponding $Q$-value, and the value of the isotope-dependent theory uncertainty $\Delta \delta_R'$ entering Eq.~(\ref{eq:ftpred_vs_ftexp}). }
\label{tab:beta_decay}
\end{table}

As mentioned in the introduction, the combined analysis of superallowed $\beta$-decay and various pion and kaon decay processes allows for the simultaneous extraction of $V_{ud}$ and $V_{us}$. 
The latest (2026) PDG average of these two elements indicates a tension with the unitarity of the CKM matrix, the so-called Cabibbo Angle Anomaly (CAA). 
Motivated by the CAA, there has been an intense interest in scrutinizing the calculations of the RC (and their associated uncertainty estimates) relevant for superallowed $\beta$-decay~\cite{Cirigliano:2024msg,Cirigliano:2024rfk,Gennari:2024sbn, Crosas:2025xyv,Cao:2025zxs,Piarulli:2026abq}. 
In particular, there are now {\it ab initio} calculations of $\delta_{\rm NS}$ from chiral EFT, which  are currently limited to the relatively light isotopes ${}^{10}$C and ${}^{14}$O but heavier systems are now being targeted. 
Within this framework, a detailed calculation of RC for ${}^{14}$O has been performed, finding good agreement with Towner and Hardy for the central value but with larger theoretical uncertainties.
These arise mainly from short-distance contributions~\cite{Cirigliano:2024msg,Cirigliano:2024rfk} that were omitted in the original calculations \cite{Towner:1992xm}. 

Taking into account this result, we perform in Sect.~\ref{subsec:stability} alternative SMEFT fits where the uncertainty associated with the nuisance parameter $\eta_2$ in Eq.~(\ref{eq:ftpred_vs_ftexp}) is increased from Eq.~(\ref{eq:delta_NS_1}) to 
\begin{equation}
\label{eq:deltaNS_inflated}
\Delta \delta_{\rm NS}^{(0)} = 9.2\cdot 10^{-4}\,.
\end{equation}
This value is chosen such that using Eq.~\eqref{eq:ftpred_vs_ftexp} reproduces the extraction of $V_{ud}$ from ${}^{14}$O performed in~\cite{Cirigliano:2024msg,Cirigliano:2024rfk}. 

Finally, we expect theoretical calculations of $\beta$-decays to significantly improve in the upcoming years, with the ultimate target being to reduce the uncertainty in RC. 
We therefore present also in Sect.~\ref{subsec:stability} alternative fits based on the optimistic scenario where the RC uncertainties are reduced such that we can extract $\widetilde V_{ud}$ from the superallowed $\beta$-decay data at $\mathcal O(10^{-4})$ uncertainty. This requires reducing the various uncertainties to 
\begin{align}\label{eq:deltaNS_reduced}
\Delta_R^V= 0.02471 \pm 0.0001\,,\qquad \Delta \delta_{\rm NS}^{(0)} =  10^{-4}\,,\qquad \Delta\delta_{\rm NS,E}^i=0.25\cdot10^{-4}\,Q_i/\mathrm{MeV}\,.
\end{align}
In this way, together with the scenario defined by Eq.~(\ref{eq:deltaNS_inflated}), we assess the stability of our results with respect to current estimates of the RC uncertainties entering Eq.~(\ref{eq:isotope-dep-corrections}), both in optimistic and in pessimistic scenarios.

\paragraph{Electroweak precision observables.}
Stringent constraints on the operators that modify the electroweak couplings of leptons and light quarks are provided by the electroweak precision observables (EWPOs) measured at the $Z$-pole by LEP and SLD~\cite{ALEPH:2005ab}.
Here we adopt the same set of pseudo-observables and associated EFT implementation as in~\cite{Celada:2024mcf,Armadillo:2026mvp}, to which we refer for the complete list of inputs and for the details of their theoretical treatment. 
These EWPOs comprise
the $Z$ lineshape observables (the total width $\Gamma_Z$, the hadronic pole cross
section $\sigma^0_{\rm had}$, and the ratios $R^0_\ell$, $R^0_b$, $R^0_c$), the
forward-backward and polarisation asymmetries ($A^{0,f}_{\rm FB}$, $A_f$, and the
left-right asymmetry $A_\ell$ from SLD), together with the $W$-boson width.

For the operators entering the interpretation of $\beta$-decays (see Table~\ref{tab:operator_definitions}), EWPOs play a central role.
In particular, the vertex-correction operators $\mathcal{O}^{(3)}_{\varphi\ell_1}$,
$\mathcal{O}^{(3)}_{\varphi \ell_2}$, and $\mathcal{O}^{(3)}_{\varphi q}$ shift the
$W\ell\nu$ and $Zff$ couplings and are therefore probed directly at the $Z$-pole,
while the four-lepton operator $\mathcal{O}^{(ijji)}_{\ell\ell}$, which enters the
extraction of $G_F$ from muon decay in Eq.~(\ref{eq:GFshift}), is constrained through
its effect on the electroweak input parameters.

As will be discussed in Sect.~\ref{sec:results}, EWPOs already constrain reasonably well the combinations of $c^{(3)}_{\varphi\ell}$, $c^{(3)}_{\varphi q}$, and $c'_{\ell\ell}$ that would
otherwise contaminate the determination of $V_{ud}$, while the semileptonic
operator $c^{(3)}_{q\ell}$ is left mostly unconstrained and requires information from LHC processes, as discussed next.

\paragraph{LHC data and HL-LHC projections.}
On top of the EWPOs, the  collider dataset considered in this work comprises the LHC measurements
entering the \smefit\,global analysis of~\cite{Celada:2024mcf,Armadillo:2026mvp}: inclusive and
differential Higgs production and decay rates, differential diboson production, and the top-quark
sector (inclusive and differential $t\bar t$, single-top, $t\bar t V$,
$t\bar t t\bar t$, and $t\bar t b\bar b$ production), in most cases based on the full LHC Run~II dataset.
These processes constrain a broad set of bosonic, Yukawa, and top-quark operators
and, through the RGE, feed into the running of the operators relevant at low
energies. 
In addition, following the updates in~\cite{Armadillo:2026mvp}, we also include charged-current high-mass Drell-Yan measurements from the LHC, whose tails provide  sensitivity to the semileptonic operator $\mathcal{O}^{(3)}_{q\ell}$ relevant for charged-current transitions in $\beta$-decays.

Concerning the HL-LHC, we use the projections constructed in~\cite{Armadillo:2026mvp},
obtained by extrapolating the Run~II measurements to an integrated luminosity of
$3\,\mathrm{ab}^{-1}$, with statistical uncertainties rescaled accordingly and a
projected reduction of the dominant systematic and theoretical uncertainties. 
Furthermore, dedicated projections for some differential processes are also considered, in particular for $t\bar{t}$ production and for Higgs pair production~\cite{terHoeve:2025omu}.
As we show in Sect.~\ref{sec:results}, the improved precision of the HL-LHC sharpens the
bounds on all these directions, including the semileptonic one probed by the
Drell-Yan tails, though the residual sensitivity is still not sufficient to lift the degeneracy affecting the extraction of $V_{ud}$.

\paragraph{FCC-ee projections.}
The projections for the FCC-ee used in this work are taken from the most recent {\sc\small SMEFiT} study of
future-collider sensitivity~\cite{Armadillo:2026mvp}, to which we refer for a
detailed description of the running scenarios, the assumed integrated luminosities,
and the theoretical treatment of each observable; here we summarise the inputs most
relevant for the low-energy analysis. 

Following the baseline FCC-ee
programme~\cite{FCC:2025lpp,FCC:2025uan}, we include the four main runs spanning
$\sqrt{s}=91$~GeV to $\sqrt{s}=365$~GeV: the $Z$-pole run (Tera-Z), the $W^+W^-$
threshold region, the $ZH$ maximum at $\sqrt{s}=240$~GeV, and the $t\bar t$
threshold run at $\sqrt{s}=365$~GeV. 
The corresponding observables are the $Z$-pole
EWPOs measured with Tera-$Z$ statistics, fermion-pair production $e^+e^-\to f\bar f$,
Higgs-boson rates, and diboson and top-quark production, with in particular optimal observables
employed in the $W^+W^-$ and $t\bar t$ channels.
See also~\cite{Celada:2026ite} for an update of the analysis of~\cite{Armadillo:2026mvp} based on descoped and upscoped scenarios for the FCC-ee running, which nevertheless do not affect any of the conclusions found in this work.

Two features of these FCC-ee projections are particularly relevant for the present work.
First, the
Tera-$Z$ run improves the determination of the vertex-correction operators
$c^{(3)}_{\varphi\ell}$  and $c^{(3)}_{\varphi q}$, as
well as of $c'_{\ell\ell}$, by around an order of magnitude or more with respect to
LEP and SLD. Second, and most importantly, fermion-pair production above the $Z$
pole constrains the semileptonic operator $c^{(3)}_{q\ell}$ through its
contact-interaction contribution far more tightly than the high-mass Drell-Yan tails
at the (HL-)LHC, with the neutral-current measurement fixing, by $SU(2)_L$ gauge
invariance, the same operator that enters the charged-current $\beta$-decay amplitude.
As will be shown in Sect.~\ref{sec:results}, it is this specific operator combination, an improved control of the vertex corrections together with improved constraint on the semileptonic direction, that allows the
FCC-ee to disentangle $V_{ud}$ from possible New Physics contributions without any loss in precision.

In this work, we mainly follow the ``aggressive'' scenario described in~\cite{deBlas:2025gyz,Armadillo:2026mvp} for the FCC-ee projections, assuming substantial improvement of theoretical uncertainties stemming from advances in higher-order calculations. The impact of reduced precision is discussed in Sect.~\ref{subsec:stability}.

%% file: sec-methodology.tex
\section{Analysis methodology}
\label{sec:methodology}

Here we describe the methodology that is used to constrain simultaneously $V_{ud}$ and the SMEFT Wilson coefficients from a combined low-energy and collider likelihood.
An important novelty of our analysis is the implementation in \smefit of extended functionalities enabling the simultaneous determination of SM parameters together with the EFT coefficients, as is already possible in other related fitting packages such as {\sc\small HEPfit}~\cite{DeBlas:2019ehy}.

\paragraph{Simultaneous SM  parameter fitting in \smefit.}
\label{eq:SM_parameter_fitting}

Here we present the first simultaneous joint determination of a SM parameter (namely $V_{ud}$) with the SMEFT Wilson coefficients within the \smefit framework. 
This determination has been performed under the assumption that the collider observables entering the fit (LEP, LHC, HL-LHC, FCC-ee) depend only weakly on $V_{ud}$, making it sufficient to only consider its impact on the superallowed $\beta$-decay predictions.
Indeed, since collider measurements do not have light-quark flavour tagging capabilities, they are essentially insensitive to the value of $V_{ud}$ and hence variations of its value can be neglected in the corresponding theoretical predictions. 

Therefore, for this specific application, $V_{ud}$ can be implemented as an additional degree of freedom only entering the low-energy likelihood, assuming an agnostic flat prior, and determined from the data through the exact evaluation of the low-energy likelihood presented in Sect.~\ref{sec:observables} without the need of any approximation such as linearisation.
For other SM parameters, such as $\alpha_s(m_Z)$ or $\alpha_{\rm EM}$, this strict separation between the SM and EFT likelihoods may not be possible.
In such cases, one may treat SM parameters in the same footing as WC, namely expanding observables as $\sigma({\boldsymbol g})=\sigma({\boldsymbol g}^{(0)}) + {\boldsymbol \kappa}^{g}_\sigma \delta {\boldsymbol g} + \ldots $ around reference values of the SM parameters ${\boldsymbol g}^{(0)}$ and then fitting the deviations $\delta {\boldsymbol g}$ as the code does with the WCs. 
We plan to systematize this feature in \smefit in future work.

Furthermore, the three nuisance parameters $\eta _i$ appearing in Eq.~(\ref{eq:ftpred_vs_ftexp}) also enter only in the analytical $\beta$-decay likelihood. 
To prevent a degeneracy with the extraction of $V_{ud}$, it is necessary to assume an informed prior, which we choose to be an univariate Gaussian centred at zero, following the method of~\cite{Falkowski:2020pma}. 
We study in Sect.~\ref{subsec:stability} the dependence of our results with respect to the assumed value of the hadronic theory error associated to these nuisance parameters. 

\paragraph{Theory calculations.}
Theoretical calculations of superallowed $\beta$-decay lifetimes have been discussed in Sect.~\ref{sec:observables}.
Those for collider observables follow the settings described in~\cite{Armadillo:2026mvp}. 
In particular, theoretical predictions in the SMEFT of LHC and FCC-ee observables are available at both linear and quadratic level.
In this work we present results only based on linear EFT calculations, since it has been demonstrated~\cite{Armadillo:2026mvp} that quadratic effects are heavily suppressed once the FCC-ee constraints are accounted for. 
Partonic matrix elements, including decay branching ratios, are evaluated accounting for NLO QCD and electroweak corrections whenever available. 
One-loop RGE effects are included systematically in all fits, and are accounted for such that flavour structures are consistent throughout. 

\paragraph{Fisher information analysis.}
The Fisher information matrix quantifies the relative sensitivity of each of the degrees of freedom entering the fit to the various experimental inputs and theoretical constraints. 
It can be evaluated both at linear and at quadratic level in the EFT expansion. 
Following~\cite{Ethier:2021bye} here we define the Fisher information matrix $I_{ij}$ as
\begin{equation}
  I_{ij}(\bm{c}) \;=\; -\,\mathrm{E}\!\left[
  \frac{\partial^2 \ln \mathcal{L}(\bm{\sigma}_{\rm exp}|\bm{c})}{\partial c_i\, \partial c_j}
  \right] , \qquad i,j = 1,\ldots,n_{\rm op} ,
  \label{eq:fisher}
\end{equation}
where $\mathrm{E}[\,]$ indicates the expectation value over the experimental data, and $\mathcal{L}(\bm{\sigma}_{\rm exp}|\bm{c})$ is the combined low-energy and collider likelihood, possibly including also theory constraints.
The
covariance matrix in the EFT parameter space, $C_{ij}(\bm{c})$, is then bounded (Cram\'er-Rao bound) by the
Fisher matrix:
$
  C_{ij} \;\geq\; \left(I^{-1}\right)_{ij}
$.

Previous implementations of the Fisher information matrix Eq.~(\ref{eq:fisher}) in {\sc\small SMEFiT} assumed a multi-Gaussian likelihood with a quadratic dependence on the WCs, which is hence not applicable to the case in which we also fit SM parameters. 
To bypass this limitation, now  the Fisher matrix is evaluated numerically for an arbitrary dependence of the likelihood on the PoI, while the best-fit values at which  Eq.~(\ref{eq:fisher}) is to be evaluated are identified by means of a gradient descent algorithm.
In this way $  I_{ij}(\bm{c})$ can be evaluated for SM parameters as well as the possible theory constraints (nuisances) entering the likelihood definition, without any assumption on their functional dependence in the global likelihood. 

\paragraph{Analysis strategy.}
The starting point of our analysis is defining which SMEFT operators are active at $\mu_0$, the starting scale for the RGE. 
This is also the scale at which the matching to UV models would take place.
In this work we always take $\mu_0 = 10$ TeV, assumed to coincide with the New Physics scale $\Lambda$, and consider that the only non-zero EFT coefficients at $\mu_0$ are, for the fits based on the $U(3)^5$ flavour structure, 
\begin{align}
    \{c^{(3)}_{q \ell}(\mu_0), c^{(3)}_{\varphi\ell}(\mu_0), c^{(3)}_{\varphi q}(\mu_0), c^{\prime}_{\ell \ell}(\mu_0) \}  \, ,
\label{eq:SMEFT_operator_basis}
\end{align}
while for the alternative fits based on the $\mathcal{G}_F$ flavour structure we have instead
\begin{align}
    \{c^{(3)}_{q \ell}(\mu_0),
    c^{(3)}_{Q \ell}(\mu_0),c^{(3)}_{\varphi\ell}(\mu_0), c^{(3)}_{\varphi q}(\mu_0),
    c^{(3)}_{\varphi Q}(\mu_0),
    c^{\prime}_{\ell \ell}(\mu_0) \}  \, ,
\label{eq:SMEFT_operator_basis_2}
\end{align}
Unless otherwise mentioned, we show results for these SMEFT coefficients at $\mu_0=10$ TeV.
In each case, we ensure that RGE evolution produces results which do not break the assumed flavour structures. 

Subsequently, the SMEFT RGEs (with {\tt wilson}) are used to evolve these four operators from $\mu_0=10$~TeV down to the scales $\mu$ at which the collider observables are provided~\cite{terHoeve:2025gey}. 
Note that due to operator mixing, at scales $\mu \le \mu_0$ many other operators in addition to those listed in Eqns.~(\ref{eq:SMEFT_operator_basis})--(\ref{eq:SMEFT_operator_basis_2}) will be generated and are accounted for in the corresponding theory predictions.
Simultaneously, the same EFT coefficients of Eq.~(\ref{eq:SMEFT_operator_basis}) are evolved down to $\mu=m_Z$ using {\tt rgevolve}~\cite{Smolkovic:2026cba}, matched to the LEFT using Eq.~(\ref{eq:smeft_combination_lprime}), and then further evolved also with {\tt rgevolve} down to $\mu\sim 2$ GeV, where they are used to evaluate the low-energy likelihood of superallowed $\beta$-decays defined in Sect.~\ref{sec:observables}.
The evaluation of the low-energy likelihood depends also on $V_{ud}$ as well as on the theory nuisance parameters.
We  have benchmarked the RGE running between {\tt rgevolve} and {\tt wilson}, finding excellent agreement. 

By means of this analysis pipeline, first we use this combined likelihood under the SM assumption to extract $V_{ud}$ and the associated theory nuisances, showing that the PDG average is correctly reproduced.
Then we run the joint SM+SMEFT fit, where $V_{ud}$ is constrained at the same time as the Wilson coefficients of Eq.~(\ref{eq:SMEFT_operator_basis}). We do this for different dataset combinations: only low-energy, then adding LHC, HL-LHC, and finally adding the FCC-ee projections.
In each case, we derive the full posterior distributions for $V_{ud}$, the SMEFT coefficients, and the theory nuisance parameters. 
We then study the stability of the results with respect to settings of the theory calculation, such as the choice of flavour assumptions, the scenarios for theoretical uncertainties at the FCC-ee, and the scenarios for the RC uncertainties entering the description of $\beta$-decays in the low-energy likelihood.

%% file: sec-results.tex
\section{Results}
\label{sec:results}

Here we present the main results of this work, namely a combined determination of the SMEFT Wilson coefficients from low-energy, (HL-)LHC, and FCC-ee observables together with the CKM matrix element $V_{ud}$.
First in Sect.~\ref{subsec:CKM_SM_fit} we present a stand-alone (SM-only) determination of $V_{ud}$, benchmarked with the PDG average, validating the implementation of the low-energy likelihood.
Then in Sect.~\ref{sec:interplay} we provide the joint extraction of $V_{ud}$ and the SMEFT coefficients for different input datasets, demonstrating that only the FCC-ee is able to fully break the degeneracy affecting the $V_{ud}$ determination from low-energy observables.
Finally, we study in Sect.~\ref{subsec:stability} the stability of our results with respect to the different scenarios for the hadronic uncertainties entering the interpretation of $\beta$-decays,  the choice of flavour structures, and the projected precision of theory uncertainties for the FCC-ee observables.

\subsection{Determination of $V_{ud}$ in the SM fit}
\label{subsec:CKM_SM_fit}

To validate our implementation of the low-energy likelihood, we carry out SM fits of the CKM matrix element $V_{ud}$ and compare the results to the average value provided by the PDG 2026~\cite{ParticleDataGroup:2026mpi}.
As discussed in Sect.~\ref{sec:beta_decays_smeft}, low-energy observables alone cannot disentangle BSM contributions from $V_{ud}$.
Therefore, the results of this SM fit can be effectively understood as a determination of $\widetilde V_{ud}$ in the presence of BSM Physics. That is, we effectively extract $\widetilde V_{ud}$ defined in Eq.~\eqref{eq:VuDeff}.  

As discussed in Sect.~\ref{sec:observables}, extracting $V_{ud}$ from the phase-space corrected lifetimes $\mathcal{F} t$ requires knowledge of the hadronic radiative corrections $\Delta_R^V$, see Eq.~(\ref{eq:CV_BSM}).
Furthermore, the relationship between the observed and predicted values of $\mathcal{F}t$ are affected by isotope-dependent long-distance radiative corrections and nuclear-structure-dependent effects, Eq.~(\ref{eq:isotope-dep-corrections}), whose uncertainties are encoded in the theory nuisance parameters $\Delta \delta_R^{'i}$, $\Delta \delta_{\rm NS}^{\rm(0)}$, and $\Delta \delta^i_{\rm NS,E}$, see Eq.~(\ref{eq:ftpred_vs_ftexp}).
Therefore, when fitting $V_{ud}$, one has to also constrain at the same time $\Delta_R^V$ and the nuisances $\eta_i$ from the low-energy likelihood.  

 \begin{figure}[t]
     \centering
\includegraphics[width=0.90\linewidth]{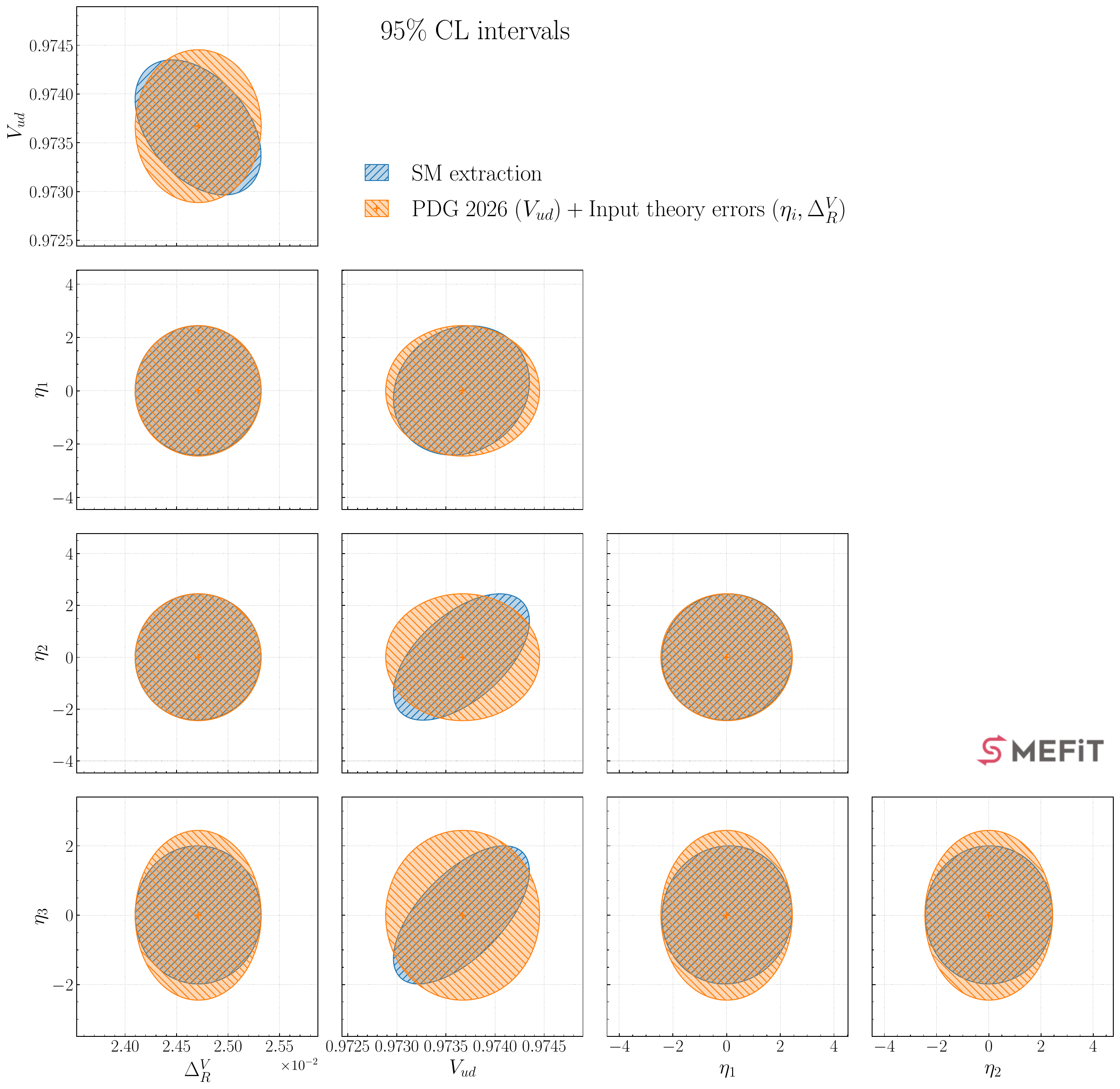}
     \caption{Results for the SM determination of $V_{ud}$ from the low-energy likelihood composed by superallowed $\beta$-decays.
     Here $V_{ud}$ is constrained alongside with the universal radiative correction $\Delta_R^V$ and the three nuisance parameters $\eta_i$ associated to the RC uncertainties in Eq.~(\ref{eq:ftpred_vs_ftexp}).
     For each pair of the five DoF  considered ($V_{ud}, \Delta_R^V$, $\eta_i$), we display the 95\% CL contours from the marginalised fit posterior distributions.
    For reference, we also indicate for each DoF pair the ellipses obtained from combining the PDG 2026 value of $V_{ud}$ with the input theory uncertainties assumed for
    $\Delta_R^V$ and the nuisances $\eta_i$ as discussed in Sect.~\ref{sec:observables}.
     }
\label{fig:SM_Vud_nuisances} 
\end{figure}

Fig.~\ref{fig:SM_Vud_nuisances}
displays the results of the {\sc\small SMEFiT} determination of $V_{ud}$ in the SM fit to the low-energy likelihood composed by superallowed $\beta$-decays, in the form of 95\% CL contours for $V_{ud}$ along with the hadronic radiative corrections $\Delta_R^V$, and the nuisance parameters $\eta_1, \eta_2,$ and $\eta_3$.
There is agreement between the resulting determination of $V_{ud}$ (which agrees with Ref.~\cite{Falkowski:2020pma}),
\be
\label{eq:vud_sm_fit}
V_{ud} = 0.97366 \pm 0.00029\, ,
\ee
and the PDG 2026 average
\be
\left| V_{ud}^{\rm (PDG)} \right| = 0.97367 \pm 0.00032 \, ,
\ee
which is based on a comparable low-energy dataset and similar, but not exactly the same, estimates of the theoretical uncertainties associated to nuclear and hadronic effects.
This agreement validates the implementation of the low-energy likelihood from superallowed beta decays in \smefit and further confirms that the latter dominates the $V_{ud}$ extraction. 

From Fig.~\ref{fig:SM_Vud_nuisances}, one also observes that our fit reproduces the pre-fit constraints imposed on $\Delta_R^V$ and $\eta_2$, which is expected by construction since these are isotope-independent nuisance parameters.
Furthermore, we find that the low-energy likelihood does not have any effect on the $\eta_1$ nuisance, but it does help to constrain $\eta_3$, the nuisance associated to the nuclear structure theory uncertainty $\Delta \delta_{\rm NS, E}^i$, by around 20\% better than the prior pre-fit assumption.
In all cases, the best-fit results coincide with the central value of the pre-fit distributions.
We have verified that the same pattern for $\Delta_R^V$ and $\eta_1, \eta_2,\eta_3$ is repeated irrespective of the input dataset and other variations in the fit settings, such as the flavour structure, and hence in the remainder of this section it is not considered further.

 \begin{figure}[t]
     \centering
\includegraphics[width=0.6\linewidth]{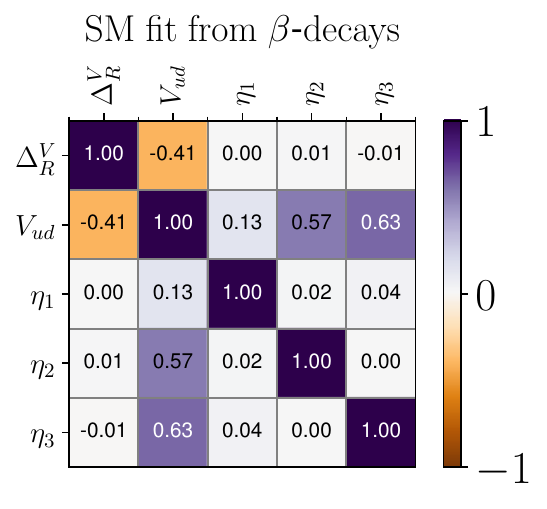}
     \caption{Correlation coefficients evaluated among the DoF entering the fit in the SM determination of $V_{ud}$ and the associated nuisance parameters from $\beta$-decays, see also Fig.~\ref{fig:SM_Vud_nuisances} for the corresponding 2D contours.
     }
\label{fig:SM_Vud_nuisances_corr} 
\end{figure}

Figure~\ref{fig:SM_Vud_nuisances_corr} displays the correlation coefficients evaluated among the DoF entering the fit in the SM determination of $V_{ud}$, whose corresponding two-dimensional contours were shown in Fig.~\ref{fig:SM_Vud_nuisances}.
The main result is that the radiative correction $\Delta_R^V$ and the nuisances $\eta_i$ are essentially uncorrelated among themselves, which means that from the fit point of view they act independently. 
Concerning $V_{ud}$, it is anti-correlated with $\Delta_R^V$, which can be understood from Eq.~(\ref{eq:CV_BSM}), and positively correlated with $\eta_2$ and $\eta_3$, which again is expected from the structure of Eq.~(\ref{eq:ftpred_vs_ftexp}).
The correlation coefficient between $V_{ud}$ and $\eta_1$ (the nuisance associated with $\Delta \delta_R^{\prime i}$) indicates that at the current level of precision, this source of theory error can be neglected.

\subsection{Combined SMEFT analysis}
\label{sec:interplay}

Next, we study the interplay between low-energy and collider constraints for the simultaneous determination of the SMEFT Wilson coefficients and the CKM matrix element $V_{ud}$.
As discussed in Sect.~\ref{sec:beta_decays_smeft}, we assume the $U(3)^5$ flavour structure as baseline and relax this assumption to the $\mathcal{G}_F$ group in Sect.~\ref{subsec:stability}. 
$\beta$-decay measurements on their own cannot disentangle shifts in $V_{ud}$ from nonzero values of the LEC $L'\equiv L'^{(1111)}$. 
The solution is provided by measurements at LEP and the LHC, and in the future at HL-LHC and FCC-ee, which constrain the four individual Wilson coefficients in Eq.~(\ref{eq:operators_smeft_matching_mz}) that match to $L'$ in Eq.~\eqref{eq:LVLL}. 
That is, while a pure low-energy SMEFT fit results in a runaway flat direction for the joint extraction of $V_{ud}$ and $L'$, collider data breaks this degeneracy.

\begin{figure}[t]
    \centering
\includegraphics[width=0.37\linewidth]{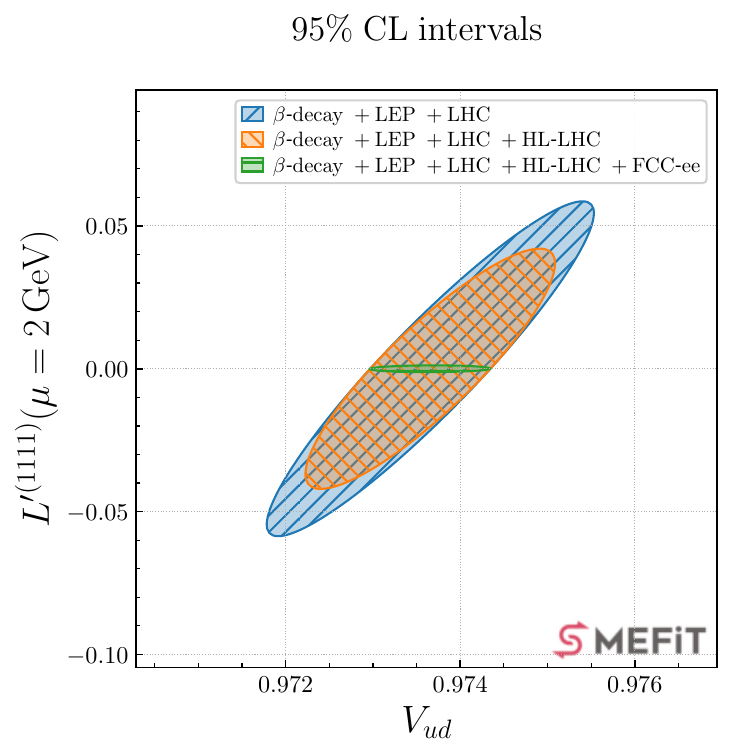}
\includegraphics[width=0.62\linewidth]{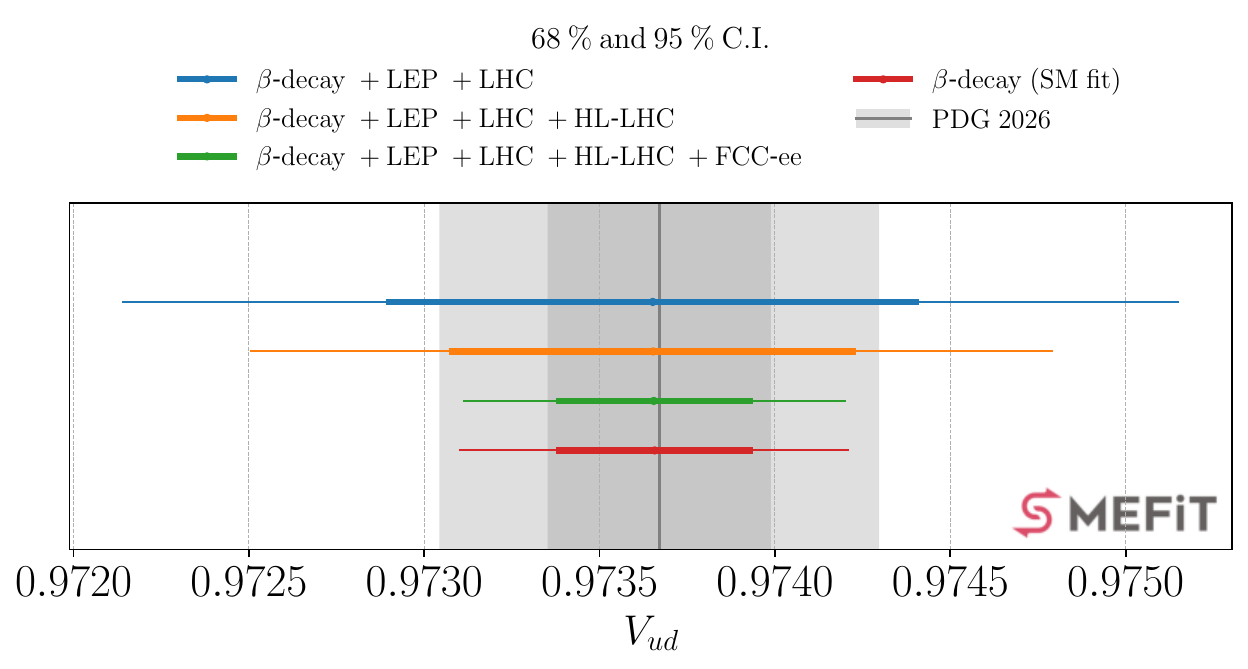}
    \caption{Left: 95\% C.I. in the SMEFT determination of the LEC $L^{\prime (1111)}(\mu = 2\,\text{GeV})$  [$\mathrm{TeV}^{-2}$] and $V_{ud}$ from fits based on different datasets, in all cases under the $U(3)^5$ flavour assumption.
    Right: the corresponding marginalised results for $V_{ud}$, compared with the 2026 PDG value as reference, as well and the SM fit result from Fig.~\ref{fig:SM_Vud_nuisances}.
    }
\label{fig:L_Vud_U(3)^5}
\end{figure}

We show this result in Fig.~\ref{fig:L_Vud_U(3)^5} for different datasets. 
Specifically, we display the marginalised 95\% CL intervals in the $(V_{ud},L'(\mu=2\text{ GeV}))$ plane for the fit based on world data ($\beta$-decays, LEP, LHC) and then in two variants, first adding the HL-LHC projections and subsequently the FCC-ee ones.
For reference, we also indicate the bound on $V_{ud}$ obtained in the SM fit and already shown in Fig.~\ref{fig:SM_Vud_nuisances}, where it was noted that this bound essentially coincides with the 2026 PDG average.

From Fig.~\ref{fig:L_Vud_U(3)^5}, one finds that by using the existing LEP and LHC data it is already possible to simultaneously constrain $V_{ud}$ and $L'$ in the SMEFT fit, since the flat direction present in the case of low-energy-only analysis is now closed.
Here the dominant constraints are provided by the high-mass CC Drell-Yan measurements of the LHC.
Subsequently, the HL-LHC projections lead to a further reduction of the uncertainties in $V_{ud}$.
Nevertheless, we also see that even with (HL-)LHC data, the precision on $V_{ud}$ is markedly worse than what is obtained in the SM fit.
Indeed, for the fits with HL-LHC projections, the  uncertainty on the CKM parameter is $\delta V_{ud}\sim 6\times 10^{-4}$, more than two times larger than obtained in the SM fit of Fig.~\ref{fig:SM_Vud_nuisances}, indicating that the uncertainty on $V_{ud}$ is no longer dominated by the uncertainties of the RC.
If one were to just use current data without HL-LHC projections, the uncertainty grows to 
$\delta V_{ud}\sim 8\times 10^{-4}$, around three times larger than in the SM-only fit.

On the other hand, Fig.~\ref{fig:L_Vud_U(3)^5} also shows that the picture changes significantly once the FCC-ee projections are accounted for.
In this case, it becomes feasible to fit both $L'$ and $V_{ud}$ simultaneously to a much higher precision, namely at the same level as the SM determination of $V_{ud}$ from superallowed $\beta$-decay data presented in Fig.~\ref{fig:SM_Vud_nuisances} with $\delta V_{ud}= 2.9\times 10^{-4}$, Eq.~(\ref{eq:vud_sm_fit}).
In addition, the constraint on $L'$ is improved thanks to the FCC-ee projections by about a factor 30 as compared to the fit with HL-LHC projections (from $\delta L'\sim 0.04$ down to $\delta L' \sim 0.0012$). 
This result shows the striking complementarity of the FCC-ee programme with low-energy experiments to simultaneously pin down SM parameters while constraining (or identifying) New Physics. 

\begin{figure}[t]
    \centering
\includegraphics[width=0.93\linewidth]{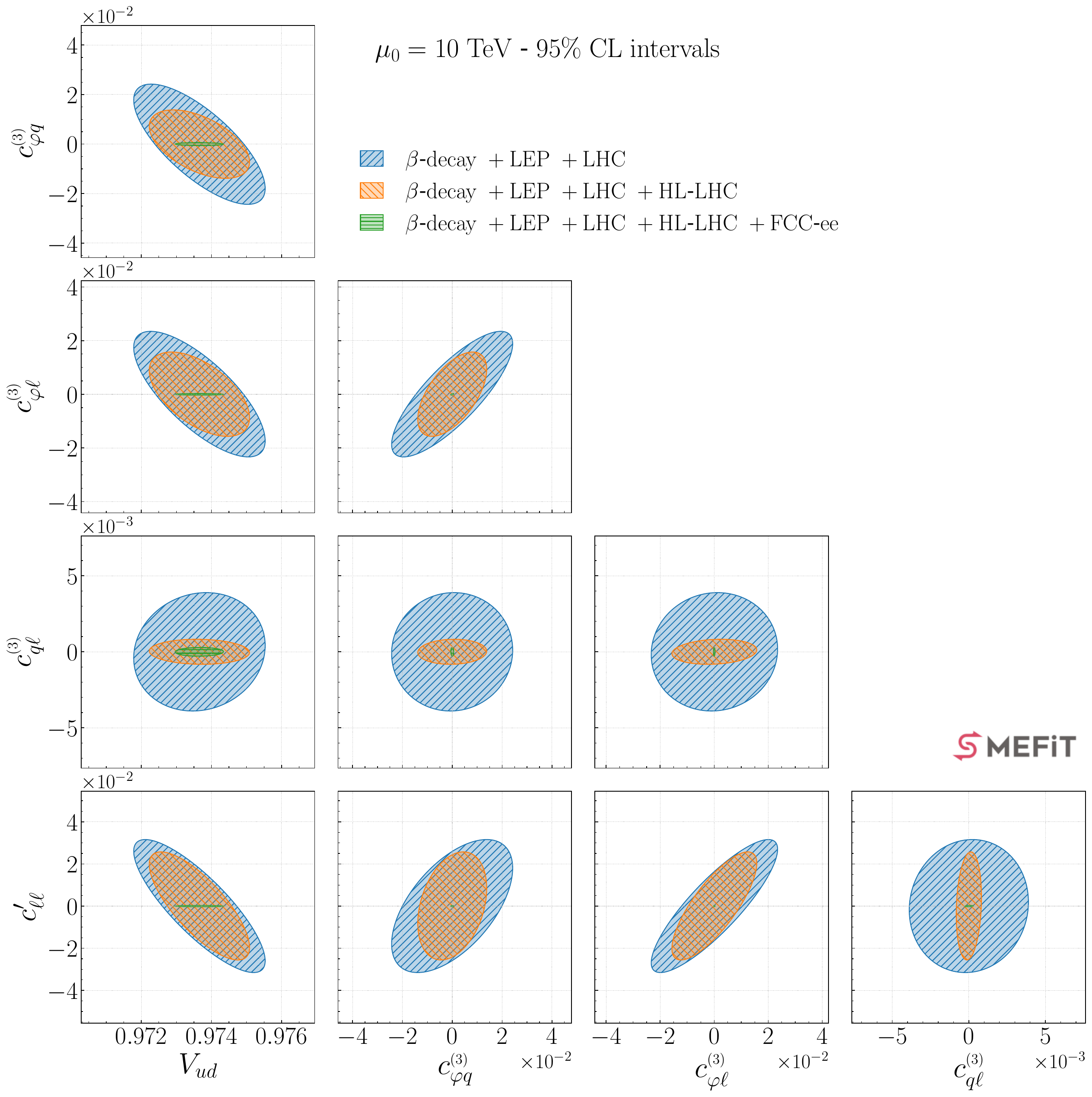}
\vspace{+0.0cm}
    \caption{Same as Fig.~\ref{fig:L_Vud_U(3)^5} now with fit results presented in terms of the four individual Wilson coefficients of Eq.~(\ref{eq:operators_smeft_matching_mz}) entering the matching relations with $L^{\prime (1111)}$.
    These WCs have units of $\mathrm{TeV}^{-2}$ and are provided at the parametrisation scale of $\mu_0=10$ TeV, rather than as $\mu=2 \,\text{GeV}$ as is the case in Fig.~\ref{fig:L_Vud_U(3)^5} (left panel) for $L^{\prime (1111)}$.}
\label{fig:SMEFT_Vud_U(3)^5}
\end{figure}

It is also interesting to present a similar comparison as that of Fig.~\ref{fig:L_Vud_U(3)^5} now at the level of the four individual Wilson coefficients of Eq.~(\ref{eq:operators_smeft_matching_mz}) entering its matching relation, which is shown in  Fig.~\ref{fig:SMEFT_Vud_U(3)^5}.
Note that Wilson coefficients are provided at $\mu_0=10$ TeV, the initial parametrisation scale, rather than as $\mu=2\text{ GeV}$ as is the case in Fig.~\ref{fig:L_Vud_U(3)^5}.
Compared to existing world data, the impact of the HL-LHC projections is particularly strong for the $c^{(3)}_{q\ell}$ two-quark-two-lepton coefficient which can be constrained by the Drell-Yan process at the  LHC~\cite{Panico:2021vav,Cirigliano:2022qdm,Allwicher:2022gkm,Greljo:2021kvv}. 
The HL-LHC enables a much better coverage of the high-energy tails where EFT effects are the largest.
However, the impact on the other three coefficients is less marked, ultimately leading to a relatively mild improvement on bounds on $V_{ud}$ from the HL-LHC extraction as compared to existing LEP and LHC world data.  

From both Figs.~\ref{fig:L_Vud_U(3)^5} and~\ref{fig:SMEFT_Vud_U(3)^5} one also observes how the impact of the FCC-ee observables is much more pronounced than that of the HL-LHC projections.
As already discussed in dedicated analyses~\cite{deBlas:2025gyz,Celada:2024mcf,Armadillo:2026mvp,Allwicher:2023shc}, the FCC-ee will greatly improve the sensitivity on electroweak-precision observables as well as on four-lepton operators, and indeed the results of Fig.~\ref{fig:SMEFT_Vud_U(3)^5} show that the resulting constraints on $c_{q\ell}^{(3)}$, $c'_{\ell\ell}$, and $c_{\varphi \ell}^{(3)}$ improve significantly as compared to the HL-LHC reference (a zoomed in version is given in Fig.~\ref{fig:corner_Vud_Lprime_SMEFiT_vs_U(3)^5}).
Furthermore, we note that the FCC-ee  completely breaks the degeneracy between $V_{ud}$ and the Wilson coefficients, leading to a determination of the former which is entirely limited by the hadronic and nuclear uncertainties affecting the interpretation of $\beta$-decays rather than by possible New Physics effects. 

Triggered by this last observation, in Sect.~\ref{subsec:stability} we study whether the main result of this work, namely the FCC-ee breaking BSM degeneracies in $V_{ud}$ extractions, still holds should theory calculations of $\beta$-decay processes become much more precise.

\begin{figure}[t]
    \centering
\includegraphics[width=0.75\linewidth]{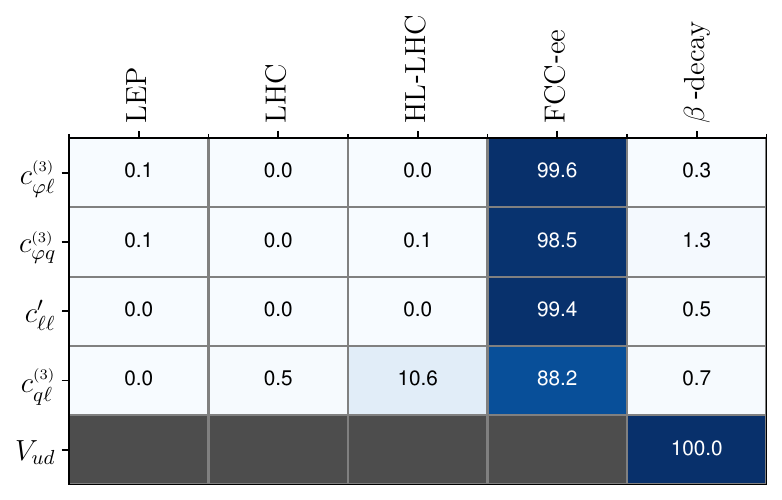}
\vspace{+0.1cm}
    \caption{The Fisher information matrix at linear order in the EFT expansion for the combined low-energy and collider likelihood.
    Each row corresponds to a DoF entering the fit, and is normalised to 100 for convention, while each column corresponds to a family of experimental datasets (see also Sect.~\ref{sec:observables}).
    }
    \label{fig:fisher-information-matrix}
\end{figure}

\paragraph{Fisher information analysis.}
Fig.~\ref{fig:fisher-information-matrix} displays the Fisher information matrix, evaluated at linear order in the EFT expansion, for the combined low-energy and collider likelihood.
Each row corresponds to a DoF entering the fit, and is normalised to 100 for convention, while each column correspond to a separate class of experimental datasets: LEP, LHC, HL-LHC, FCC-ee, and $\beta$-decays. 
We recall that here we only consider the diagonal entries of the matrix, and that off-diagonal correlations are potentially large and have implications for the fit results.

From Fig.~\ref{fig:fisher-information-matrix} one observes that the FCC-ee projections completely dominate the sensitivity on all four Wilson coefficients $c_{\varphi \ell}^{(3)}$, $c_{\varphi q}^{(3)}$,   $c'_{\ell\ell}$, and $c^{(3)}_{q\ell}$.
The additional information provided by the other datasets, including the low-energy likelihood from $\beta$-decays, is in all cases much smaller.
Only the HL-LHC provides somewhat competitive information on the two-lepton-two-quark operators $c^{(3)}_{q\ell}$, predominantly through high-mass Drell-Yan production.
$V_{ud}$ is, by construction, only constrained by superallowed $\beta$-decays.
Therefore, the Fisher information analysis confirms the general findings of this section: in the combined collider and low-energy likelihood, FCC-ee fixes the New Physics directions enabling a clean extraction of $V_{ud}$ from $\beta$-decays.

\begin{figure}[t]
   \centering
\includegraphics[width=0.8\linewidth]{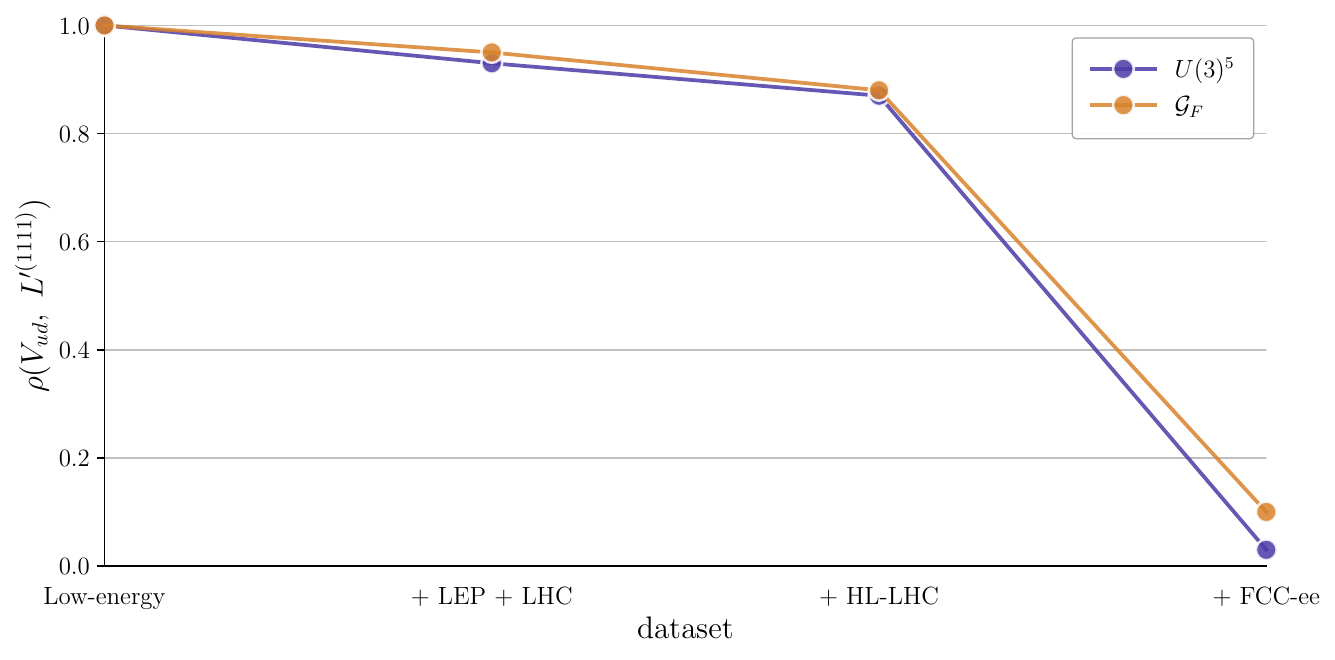}
   \caption{Dependence of  the correlation coefficient $\rho$ evaluated between $V_{ud}$ and $L'(\mu=2 \ {\rm GeV})$ as a function of the input dataset, for the two flavour structures that we consider in this work. 
A value $\rho \sim 1$ indicates a degeneracy in the ($V_{ud},L'$) space and hence a flat direction, while instead $\rho \sim 0$ indicates that the two parameters are decoupled and can be separately constrained from a joint fit.
}
\label{fig:correlation_evolution}
\end{figure}

\paragraph{Correlation analysis.}
Fig.~\ref{fig:correlation_evolution} displays the dependence of the correlation coefficient $\rho$ evaluated between $V_{ud}$ and $L^\prime(\mu=2~{\rm GeV})$ as a function of the input datasets. 
A value $\rho \sim 1$ indicates a degeneracy in the ($V_{ud},L^\prime$) space and hence a flat direction, while instead $\rho \sim 0$ indicates that the two parameters are decoupled and can be robustly determined in the context of a joint fit.
This correlation coefficient is provided by our baseline fits, obtained with the $U(3)^5$ flavour structure, and for those based on the alternative $\mathcal{G}_F$ group, to be further studied in Sect.~\ref{subsec:stability}.
From this correlation analysis we see that, for both flavour assumptions, the HL-LHC breaks the flat direction between $V_{ud}$ and $L^\prime$ present from $\beta$-decays but a partial degeneracy is still present.
Instead, once the FCC-ee constraints are accounted for, the correlation coefficient essentially vanishes, indicating that the two DoF become decoupled and can be separately determined from a joint fit.

\subsection{Stability analysis}
\label{subsec:stability}


We now study the stability of the results presented in Sect.~\ref{sec:interplay} with respect to variations of the theoretical inputs entering the analysis.
First, we study the dependence of the results with respect to the estimates of the nuclear and hadronic uncertainties entering the interpretation of $\beta$-decays discussed in Sect.~\ref{sec:observables}.
Specifically, we consider a scenario with a conservative estimate of the isotope-dependent nuclear structure correction $\delta_{\rm NS}$ in Eq.~(\ref{eq:isotope-dep-corrections}), as well as an optimistic scenario where these theory uncertainties are markedly decreased. 
Second, we study the dependence of the fit with respect to the flavour structure, by comparing the baseline $U(3)^5$ with the more general $\mathcal{G}_F$ symmetry group.
Third, we quantify how our results are modified for different scenarios concerning the projected theory predictions entering the FCC-ee observables.

\paragraph{Impact of hadronic uncertainties estimates.}
As discussed in Sect.~\ref{sec:observables}, the interpretation of superallowed $\beta$-decays in terms of CKM matrix elements and  New Physics effects relies heavily on the estimate of the various relevant hadronic uncertainties, in particular the universal radiative correction $\Delta_R^V$ in Eq.~(\ref{eq:CV_BSM}) and the three hadronic nuisance parameters entering Eq.~(\ref{eq:ftpred_vs_ftexp}). Here we quantify the role played by hadronic uncertainties in our results, in particular by presenting fits based on a conservative and an optimistic estimate of the theory nuisance parameter $\Delta\delta_{\rm NS}^{(0)}$.

\begin{figure}[ht!]
    \centering
    \includegraphics[width=0.99\linewidth]{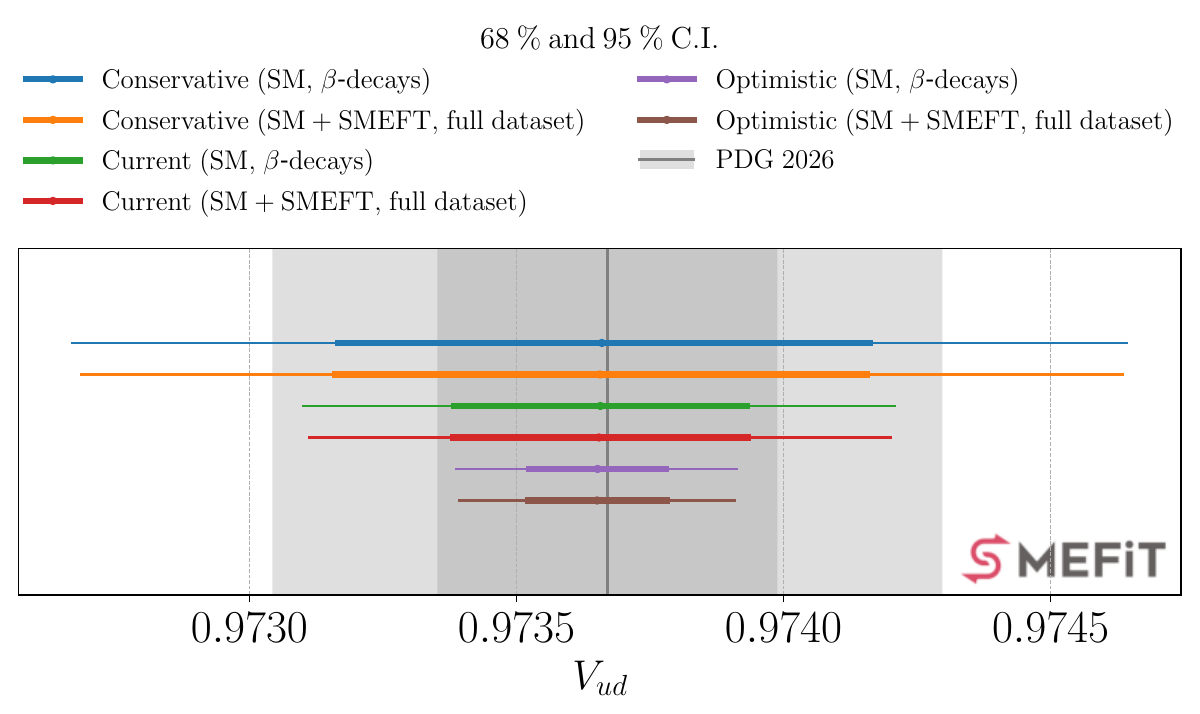}
    \caption{Comparison of $V_{ud}$ as determined in the SM and SMEFT fits based on a more conservative estimate of the theory error on the
    radiative uncertainty  $\Delta\delta_{\rm NS}^{(0)}$, Eq.~(\ref{eq:deltaNS_inflated}) (uppermost pair) with respect to the baseline choice of Eq.~(\ref{eq:delta_NS_1}) (middle pair), 
    as well as for the optimistic scenario defined 
    by Eq.~(\ref{eq:deltaNS_reduced}) (lowermost pair) for future theory uncertainties in $\beta$-decay calculations.
    In all cases, results are obtained with the $U(3)^5$ flavour structure.
    For reference, the PDG 2026 average of $V_{ud}$ is also indicated with a grey band.
    The inner and thicker (outer and thinner) bars indicate the 68\% (95\%) C.I. in each case.
    \label{fig:vud_comparison}
    }
\end{figure}

For the fits presented in Sect.~\ref{sec:interplay} we use as default the value $\Delta\delta_{\rm NS}^{(0)} = 3.3 \cdot 10^{-4}$ quoted in~\cite{Hardy:2020qwl,Seng:2018qru}.
This estimate has been argued to be optimistic, and here we investigate a more conservative value of $\Delta\delta_{\rm NS}^{(0)} = 9.2 \cdot 10^{-4}$ inspired by the chiral EFT analysis of~\cite{Cirigliano:2024msg}.
We also consider a scenario where theoretical control on the radiative corrections entering $\beta$-decays is significantly improved, see Eq.~\eqref{eq:deltaNS_reduced}.
Figure~\ref{fig:vud_comparison} compares the results for $V_{ud}$ in the SM and SMEFT fits for the ``conservative'' and ``optimistic'' scenarios, given by the uncertainties in Eqs.~\eqref{eq:deltaNS_inflated} and \eqref{eq:deltaNS_reduced} respectively, to the ``current'' scenario with the values of uncertainties used in the rest of this work. The results are obtained in all cases under the $U(3)^5$ baseline flavour structure, and with the complete dataset considered in this work. 
The analysis shows that the FCC-ee constraints are sufficient to maintain the SM-fit precision on $V_{ud}$ even in the optimistic scenario for RC uncertainties, where under the SM assumption the uncertainty of $V_{ud}$ reaches the $10^{-4}$ level.
Therefore, we conclude that once the FCC-ee data become available, the precision expected for the $V_{ud}$ extraction from $\beta$-decays will be entirely limited by the low-energy theory and experimental uncertainties.

\begin{figure}[t]
    \centering
\includegraphics[width=0.9\linewidth]{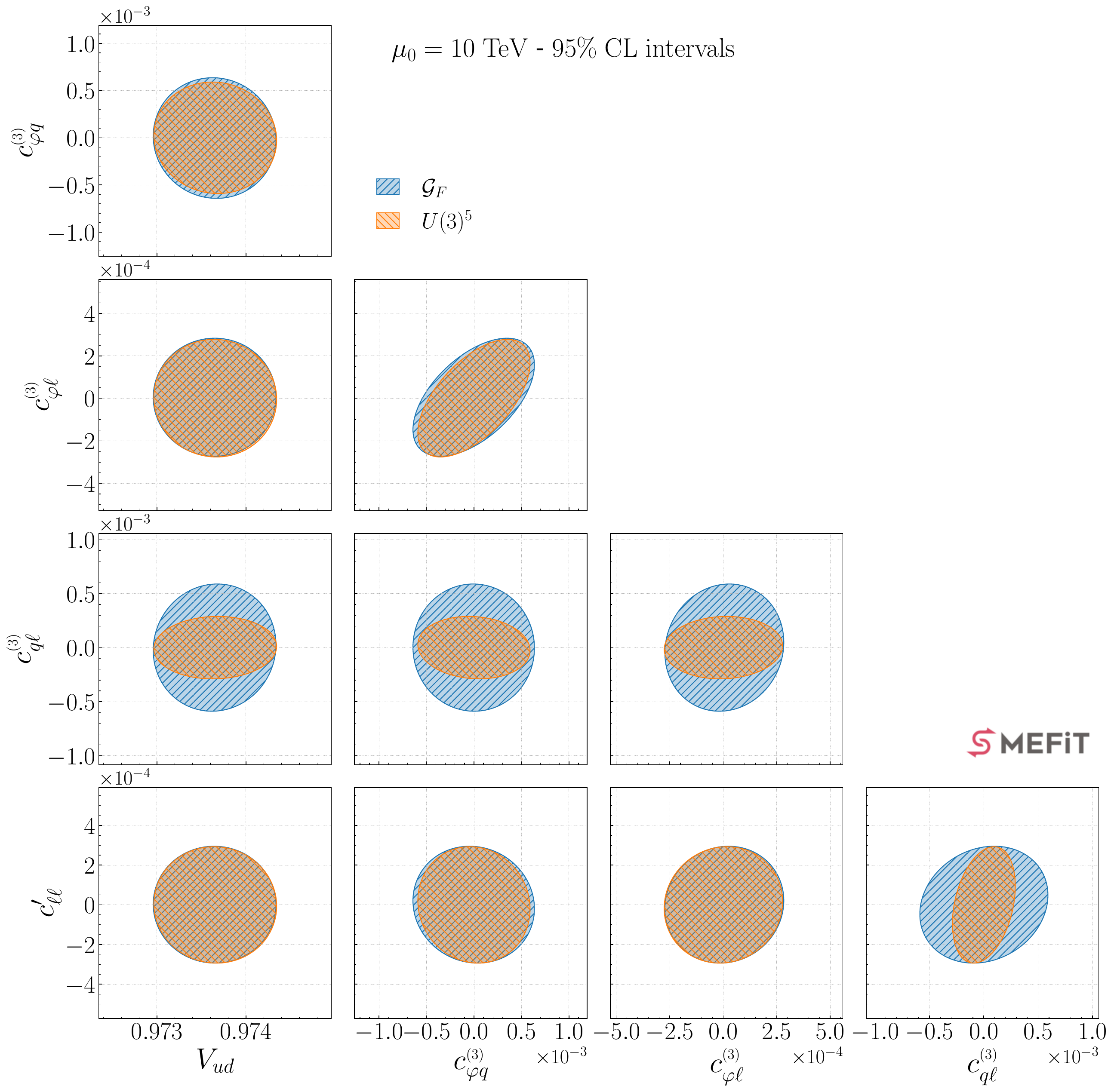}
    \caption{Same as Fig.~\ref{fig:SMEFT_Vud_U(3)^5} for the SMEFT fits containing the full dataset (LEP + LHC + HL-LHC+ FCC-ee+$\beta$-decays), comparing the results obtained with the $U(3)^5$ (baseline) and with  the $\mathcal{G}_F$ (alternative) flavour structures.
   } \label{fig:corner_Vud_Lprime_SMEFiT_vs_U(3)^5}
\end{figure}

\paragraph{Impact of the flavour structure.}
Fig.~\ref{fig:corner_Vud_Lprime_SMEFiT_vs_U(3)^5} presents the same results as Fig.~\ref{fig:SMEFT_Vud_U(3)^5} for the SMEFT fits containing the FCC-ee projections, now comparing the results obtained with the $U(3)^5$ (baseline) and with the $\mathcal{G}_F$ (alternative) flavour structures.
See Table~\ref{tab:SMEFiT_and_U(3)^5} for a summary of the WCs which are activated in each of the two scenarios. 
 While the two sets of flavour assumptions result in identical predictions for $\beta$-decays, they differ for collider observables, in particular concerning processes involving top quarks, and hence they assess the stability of our analysis with respect to the flavour structures entering the interpretation of the LHC and FCC-ee data.
 Specifically, $c_{q\ell}^{(3)}\ne c_{Q\ell}^{(3)}$ and $c_{\varphi q}^{(3)}\ne c_{\varphi Q}^{(3)}$ in the  $\mathcal{G}_F$ scenario, as opposed to the flavour universal $U(3)^5$ where these WC are set to be equal.

From Fig.~\ref{fig:corner_Vud_Lprime_SMEFiT_vs_U(3)^5} we find that the marginalised bounds on $V_{ud}$ are essentially identical in the two flavour scenarios, confirming the robustness of the main finding of Sect.~\ref{sec:interplay}. 
The two additional Wilson coefficients which are fitted in the $\mathcal{G}_F$ case lead to somewhat  weaker constraints on the semileptonic coefficient $c^{(3)}_{q\ell}$ without weakening the $V_{ud}$ extraction. 
It would be interesting to assess how this picture is modified should more WCs become activated as DoF at $\mu_0 = 10$ TeV also in the context of more general flavour structures; we plan to study this in future work. 

\paragraph{Impact of theoretical uncertainties in FCC-ee projections.}
The FCC-ee projections considered in this work inherit the treatment of SM theoretical
uncertainties adopted in~\cite{Armadillo:2026mvp}, which in turn follow the prescriptions of the 2026 European
Strategy Physics Briefing Book~\cite{deBlas:2025gyz} (see also the recent update of~\cite{deBlas:2026vrz}).
There, three main sources of theory uncertainties at the FCC-ee are accounted for. 
First, the uncertainties from missing higher-order (MHO) perturbative corrections to the relevant lepton-collider observables, namely Higgs production and decay, the $Z$-pole EWPOs, $\alpha_{\rm EW}$, and the $W$-boson width and branching ratios. 
Second, the theory inputs required to extract the $Z$-pole EWPOs from the measured lineshape, such as the modelling of the background subtraction.
Third, the parametric uncertainties from the propagation of the projected experimental precision on the SM input parameters, which in the $m_W$ scheme adopted here are $\{m_W, m_Z, G_F, m_H, m_t, \alpha_s\}$, and which
are included through their full theory covariance matrix~\cite{Mildner:2024wbl}.

The first two, reducible, sources of theory uncertainties are modelled according to four benchmark scenarios for the foreseen precision~\cite{deBlas:2025gyz} by the time the FCC-ee starts operations: a ``current'' scenario, in which the uncertainties remain at their present size; ``conservative'' and ``aggressive" scenarios, corresponding respectively to moderate and substantial improvements from progress in higher-order calculations; and an {\rm ideal} scenario, in which all SM theory uncertainties are taken to be negligible compared to the experimental precision. 
These four scenarios are directly relevant for our analysis, since the operators controlling the extraction of $V_{ud}$, in particular $c^{(3)}_{\varphi\ell}$, $c^{(3)}_{\varphi q}$, and $c'_{\ell\ell}$, are constrained predominantly by the Tera-$Z$ EWPOs, whose ultimate reach can be limited by SM theory rather than by statistics. 
As indicated in Sect.~\ref{sec:observables}, our baseline results have been obtained with the ``aggressive'' scenario, and here we assess how these change if the ``current" or  ``conservative" scenarios are used instead.

\begin{figure}[t]
    \centering
\includegraphics[width=0.37\linewidth]{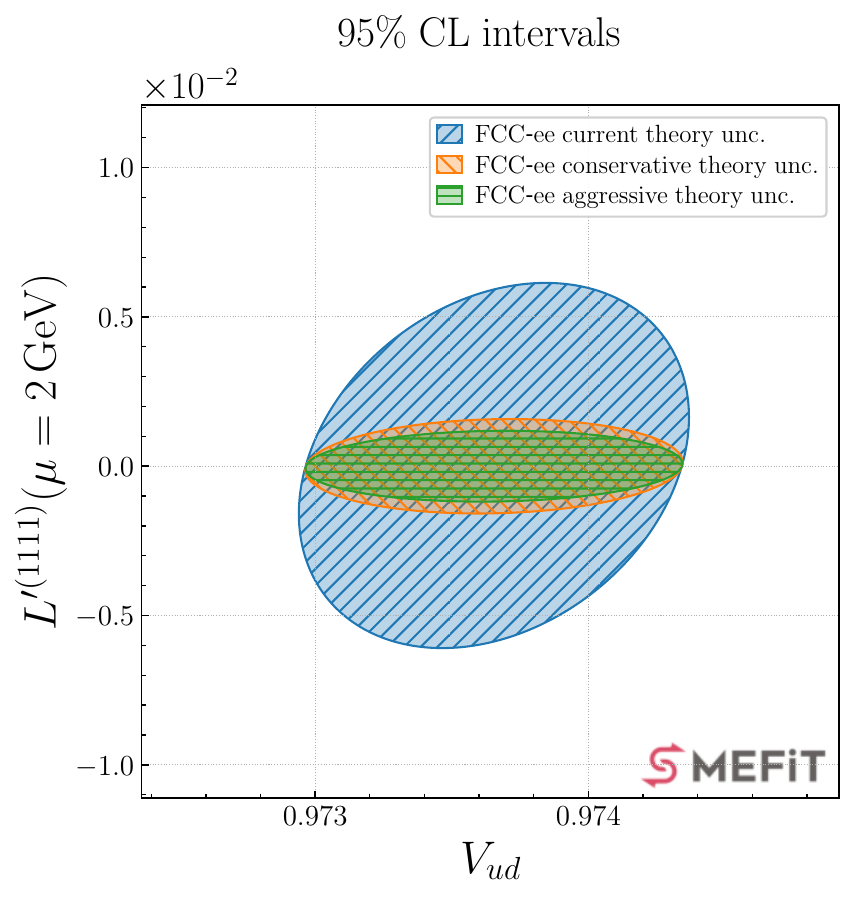}
\includegraphics[width=0.62\linewidth]{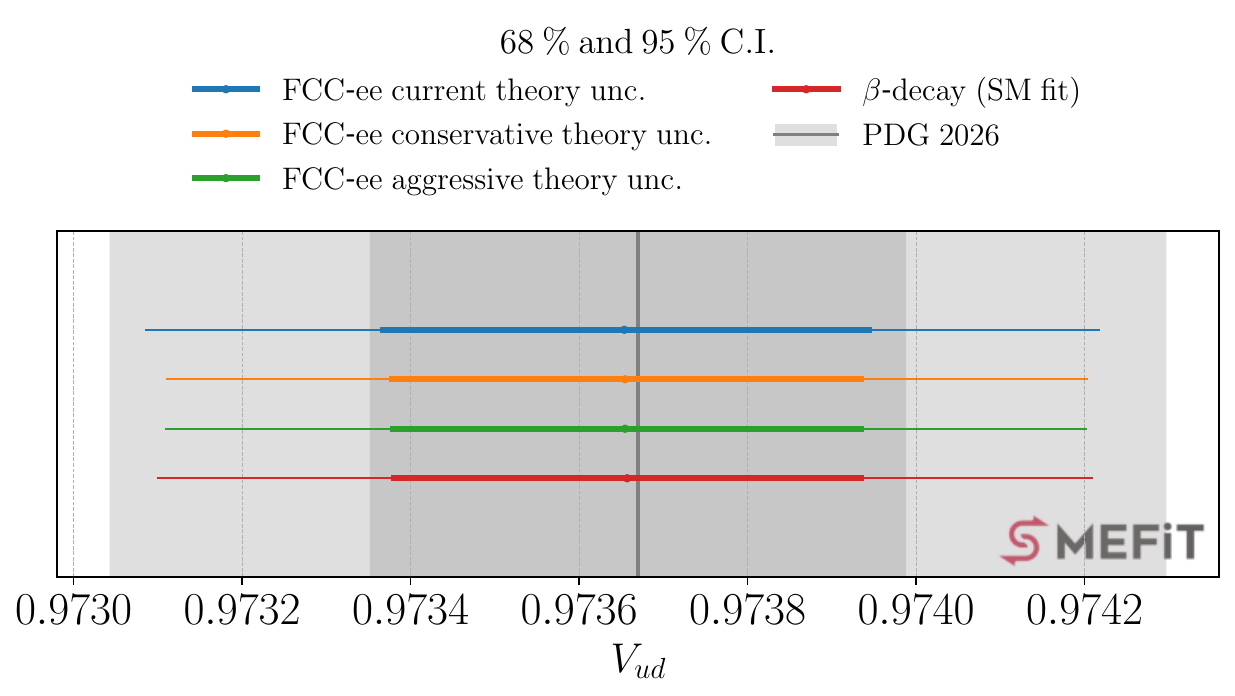}
    \caption{Left: same as the left panel of Fig.~\ref{fig:L_Vud_U(3)^5}, now comparing
    the SMEFT fits to the full dataset (LEP + LHC + HL-LHC+ FCC-ee+$\beta$-decays) in three different scenarios for the theoretical uncertainties affecting the FCC-ee predictions: the ``aggressive'' (baseline in this work),
    ``conservative'', and ``current'' scenarios.
    Right: same results, now highlighting the stability of the $V_{ud}$ determination in the same three scenarios. 
   } 
   \label{fig:FCC_tcov_scenarios}
\end{figure}

With this motivation, Fig.~\ref{fig:FCC_tcov_scenarios} presents, in the same format as Fig.~\ref{fig:L_Vud_U(3)^5}, the 95\% CL intervals in the $\lp V_{ud},L^\prime\rp$ plane comparing the results of SMEFT fits to the full dataset (LEP + LHC + HL-LHC+ FCC-ee+$\beta$-decays) in three different scenarios for the theoretical uncertainties affecting the FCC-ee predictions: aggressive (baseline choice in this work), conservative, and current.
In all three scenarios, the degeneracy between $L^\prime$ and $V_{ud}$ is broken and the precision achievable for the latter is essentially unchanged.
Furthermore, results based on the ``conservative'' and ``aggressive'' scenarios are essentially the same.
Therefore, from Fig.~\ref{fig:FCC_tcov_scenarios} we conclude that the results presented in this section are stable with respect to the projected values of theory errors entering the interpretation of FCC-ee observables.

%% file: sec-summary.tex
\section{Summary and outlook}
\label{sec:summary}

In this work we have presented a combined interpretation of superallowed $\beta$-decays and high-energy collider observables within the SMEFT.
The overarching goal was assessing to which extent the CKM matrix element $V_{ud}$ can be disentangled from possible New Physics contributions, distorting the theoretical interpretation of $\beta$-decay lifetimes, through collider constraints.
This goal has been achieved by extending the {\sc\small SMEFiT}
framework with a dedicated low-energy likelihood built from the
most precise data on superallowed nuclear $\beta$-decays, including a consistent treatment of the associated radiative, nuclear-structure, and isospin-breaking uncertainties through nuisance parameters.

This low-energy likelihood, 
available from the \href{https://github.com/LHCfitNikhef/smefit_database}{public {\sc\small SMEFiT} database},  is combined with the LHC and FCC-ee likelihoods by means of RGEs and matching to the LEFT below the electroweak scale, ensuring a consistent treatment of operator running and mixing from $\mu\sim 2$ GeV all the way up to $\mu_0 \sim 10$ TeV.
We have validated our implementation through a stand-alone SM fit of $V_{ud}$
from the superallowed $\beta$-decay dataset, finding agreement with the PDG average and confirming that a low-energy likelihood composed by superallowed $\beta$-decays dominates the $V_{ud}$ determination.

Our main finding is that current world data, as well as its extension with HL-LHC projections, cannot break the degeneracy between $V_{ud}$ and the New Physics contributions arising from low-energy observables.
One reason for this is that the semileptonic direction $c^{(3)}_{q\ell}$ is only weakly constrained by the LHC Drell-Yan data, and as a result, the uncertainty on $V_{ud}$ significantly inflates relative to the SM fit once these SMEFT operators are activated.
The inclusion of the
FCC-ee projections then pins down all operators and restores a determination of $V_{ud}$ at the same per-mille precision level as the SM fit, now cleanly separated from possible New Physics and without any appreciable loss in precision.
This result, stable upon variations of the flavour structure and an increased future control over radiative corrections, illustrates the  synergies between low-energy observables and the FCC-ee, with the higher-order quantum corrections encoded in the RGEs being key to connect information collected at different energy scales.

The present analysis is meant as a first proof of concept to quantify the connections between FCC-ee and low-energy observables and can be extended in several directions.
First, from the point of view of experimental data, we plan to include datasets such as neutron decay, mirror transitions, and pion and kaon two- and three-body decays, to enable the simultaneous determination of $V_{ud}$ and $V_{us}$ and hence provide a genuine test of first-row CKM unitarity within the SMEFT, allowing a direct assessment of the Cabibbo angle anomaly and the possible role played by New Physics in their resolution. 
It would also be interesting to replace the nuisance-parameter treatment of the Towner-Hardy radiative corrections~\cite{Falkowski:2020pma} with recent effective-field-theory and {\it ab initio} determinations of the nucleus-dependent corrections~\cite{Cirigliano:2024rfk,Cirigliano:2024msg,Gennari:2024sbn,Seng:2026hwk}, which provide a systematically improvable framework with more robust uncertainty estimates.

Second, from the SMEFT theory side, the minimal set of four (six) \smefit\,degrees of freedom considered here under $U(3)^5$ ($\mathcal{G}_F$) could be enlarged to a significantly broader subset of the SMEFT parameter space, to match {\it e.g.} the $n_{\rm op}=61$ directions constrained in~\cite{Armadillo:2026mvp}, in particular to those most relevant for the proposed resolution of the Cabibbo angle anomaly such as the right-handed Higgs current SMEFT operator $i( \bar u_R \gamma^\mu d_R)\,(\tilde \varphi^\dagger D_\mu \varphi)$ \cite{Cirigliano:2023nol}.
Along the same considerations, it may be interesting to relax our flavour structures, both $U(3)^5$ and $\mathcal{G}_F$, in order to test the robustness of our conclusions against a more general New Physics parametrisation.
This would also allow matching our results to a broader family of UV completions, in order to quantify the impact of low-energy data on the mass and coupling parameter space of explicit BSM models~\cite{terHoeve:2023pvs}.

Third, while the FCC-ee is the preferred option put forward by the European Strategy for Particle Physics, it is not the only one, and hence it would be worthwhile to revisit the analysis presented in this work to other options for future colliders studied in~\cite{Armadillo:2026mvp}, from the Linear Collider Facility (LCF)~\cite{LinearCollider:2025lya,LinearColliderVision:2025hlt} to LEP3~\cite{Anastopoulos:2025jyh} and the de-scoped FCC-ee~\cite{Celada:2026ite}, among others. 
In this way, one could assess whether our main finding, namely that the FCC-ee can disentangle $V_{ud}$ from New Physics in the analysis of superallowed $\beta$-decays, also holds for other options for future lepton colliders. 

All results presented in this work can be reproduced through  the open-source  {\sc\small SMEFiT} \href{https://smefit.science/}{code}, providing a flexible toolbox to explore the rich interplay between the physics reach of low-energy precision
measurements and that of future high-energy colliders.